\documentclass[aps,nofootinbib,amsmath,prd,showpacs,superscriptaddress,12pt,notitlepage]{revtex4-1}

\usepackage[paperwidth=210mm,paperheight=297mm,centering,hmargin=2.4cm,tmargin=2.7cm,bmargin=3.3cm]{geometry}

\usepackage{graphicx}  
\usepackage{dcolumn}   
\usepackage{bm}        
\usepackage{amssymb}   

\usepackage[utf8]{inputenc}
\usepackage{slashed}
\usepackage{epsfig}
\usepackage{amsmath,amsfonts,mathtools}  
\usepackage{color,bbm}

\usepackage[dvipsnames]{xcolor}

\usepackage{tabularx}
\newcolumntype{R}{>{\raggedleft\arraybackslash}X}

\numberwithin{equation}{section}

\makeatletter
\renewcommand{\p@subsection}{}
\renewcommand{\p@subsubsection}{}

\makeatother

\usepackage{xfrac}

\newcommand{\Z}{\mathbb{Z}}
\def\cei^#1_#2{\lower\fontdimen17\textfont2\vbox{%
   \baselineskip=\fontdimen17\textfont2 \advance\baselineskip by\fontdimen14\textfont2
   \halign{\hfil$\scriptstyle##$\hfil\cr#1\cr#2\cr}%
}}

\begin{document}



\newcommand{\colH}[1]{\textcolor{blue}{#1}} 
\newcommand{\colO}[1]{\textcolor{DarkGreen2}{#1}} 
\newcommand{\colA}[1]{\textcolor{magenta}{#1}} 
\newcommand{\coled}[1]{\textcolor{red}{(\sc #1)}}


\title{A functional perspective on emergent supersymmetry}

\author{Holger Gies}
\email{holger.gies@uni-jena.de}
\affiliation{Theoretisch-Physikalisches Institut, Friedrich-Schiller-Universit\"{a}t Jena,
Max-Wien-Platz 1, 07743 Jena, Germany}
\affiliation{Helmholtz Institute Jena,
Fröbelstieg 3, 07743 Jena, Germany}

\author{Tobias Hellwig}
\email{tobias.hellwig@uni-jena.de}
\affiliation{Theoretisch-Physikalisches Institut, Friedrich-Schiller-Universit\"{a}t Jena,
Max-Wien-Platz 1, 07743 Jena, Germany}

\author{Andreas Wipf}
\email{andreas.wipf@uni-jena.de}
\affiliation{Theoretisch-Physikalisches Institut, Friedrich-Schiller-Universit\"{a}t Jena,
Max-Wien-Platz 1, 07743 Jena, Germany}

\author{Omar Zanusso}
\email{omar.zanusso@uni-jena.de}
\affiliation{Theoretisch-Physikalisches Institut, Friedrich-Schiller-Universit\"{a}t Jena,
Max-Wien-Platz 1, 07743 Jena, Germany}

\begin{abstract}
We investigate the emergence of ${\cal N}=1$ supersymmetry in the long-range behavior 
of three-dimensional parity-symmetric Yukawa systems.
We discuss a renormalization approach that manifestly preserves supersymmetry 
whenever such symmetry is realized,
and use it to prove that supersymmetry-breaking operators are irrelevant,
thus proving that such operators are suppressed in the infrared.
All our findings are illustrated with the aid of the $\epsilon$-expansion and a functional variant of perturbation theory,
but we provide numerical estimates of critical exponents that are based on the non-perturbative functional renormalization group.
\end{abstract}

\maketitle



\section{Introduction}

Symmetries shape the structure of physical systems. In turn, the
dynamics of physical systems can influence the status and realization
of symmetries. Most prominently in quantum field theories as well as
statistical or many-body systems, fluctuations can contribute to the
spontaneous or anomalous breakdown of symmetries. In such a case, the
symmetry is present at a microscopic level, but macroscopically broken
through long-range ordering or anomalous fluctuations.

The present work investigates an inverse phenomenon where a symmetry
may not be present at the level of the microscopic interactions, but
can emerge in the long-range physics as a consequence of
fluctuations. Emergent space or spacetime symmetries are a well-known
phenomenon in solid-state physics, where atomic or molecular lattices
generically break rotational invariance microscopically, but such symmetry is
nevertheless restored for macroscopic properties. Emerging Lorentz
symmetry often forms the basis of lattice formulations of relativistic
quantum field theories \cite{Chadha:1982qq}, and is also expected at
the critical points in specific condensed matter systems such as
graphene \cite{Herbut:2006cs,Juricic:2009px,Herbut:2009vu}. The
emergence of internal symmetries has also been discussed both in
fermionic \cite{Roy:PhysRevB.84.113404} as well as bosonic
\cite{Calabrese:2002bm,Eichhorn:2013zza} systems.

Recently, the emergence of supersymmetry and thus of a nontrivial
combination of spacetime and internal symmetries has received a great
deal of attention \cite{Gomes:2015eea}. Concrete realizations in (2+1) dimensional lattice
systems have been worked out in
\cite{Lee:2006if,Grover:2013rc,Ponte:2012ru,Jian:2014pca}, and first
scenarios in (3+1) dimensions have been proposed in
\cite{Goh:2003yr,Jian:2014pca,Antipin:2011ny}. Typical constructions start with
gapless fermionic degrees of freedom; their bosonic counterparts may
also be added on a fundamental level or may arise as composite order
parameter fields. In the latter case, the bosons can naturally match
the criterion of a gapless spectrum in the vicinity of a quantum phase
transition.

This line of argument suggests (2+1)-dimensional systems with
relativistic fermion degrees of freedom as candidate systems for
emergent supersymmetry, as they exhibit quantum phase transitions at
sufficiently strong interactions \cite{Sonoda:2011qd}. These phase transitions are
reflected by interacting UV fixed points of the renormalization group
(RG). In fact, there is a substantial body of literature on such
models as they can give rise to critical phenomena where -- in
addition to the dimension and the symmetry of the (bosonic) order
parameter -- also the number and structure of the (fermionic)
long-range degrees of freedom characterize the universal
properties. Their quantitative determination has been pursued by a
variety of methods including $\epsilon$ and $1/N$ expansions
\cite{Gracey:1990sx,Hands:1992be,Vasiliev:1992wr,Rosenstein:1993zf,Moshe:2003xn,Fei:2016sgs,Gracey:2016mio,Zerf:2016fti,Mihaila:2017ble},
Monte-Carlo simulations
\cite{Hands:1992be,Karkkainen:1993ef,Christofi:2006zt,Chandrasekharan:2013aya,Wang:2014cbw,Li:2014aoa,Hesselmann:2016tvh,Chandrasekharan:2011mn,Hands:2016foa,Schmidt:2016rtz},
as well as the functional RG
\cite{Rosa:2000ju,Hofling:2002hj,Braun:2010tt,Mesterhazy:2012ei,Jakovac:2014lqa,Vacca:2015nta,Borchardt:2015rxa,Knorr:2016sfs}. These
models have recently received a great deal of attention as effective
models describing phase transitions from a disordered
(e.g., semi-metallic) to an ordered (e.g., Mott-insulating or
superconducting) phase \cite{Herbut:2006cs,Raghu:2007ger,Juricic:2009px,Herbut:2009vu,Classen:2015mar}

In the present work, we investigate the emergence of supersymmetry in
a (2+1) dimensional Yukawa-type model with a single Majorana fermion
and a dynamical real scalar order parameter field. This model may be
viewed as the simplest representative of the class of so-called chiral
Ising models; for a larger fermionic content, the latter includes the
Gross-Neveu-Yukawa models featuring the symmetries of the Gross-Neveu
model. For very particular values of the masses and couplings, our model
reduces to a supersymmetric Wess-Zumino model. In this work, we
quantitatively analyze how the supersymmetric model emerges within the
more general theory space.

On the method side, the functional RG is highly advantageous for the
present model, as it works directly in (2+1) dimensions using the
physical degrees of freedom, and is not inflicted by sign
problems. Apart from using the functional RG as a quantitative tool,
we focus on two conceptual aspects: first, we make direct contact with
the perturbative $\epsilon$ expansion technique, working on a fully
functional level. Second, we use a manifestly supersymmetric off-shell
formulation of the functional RG allowing us to investigate both the
subspace of the supersymmetric theory as well as the flow in the
nonsupersymmetric vicinity in the space of theories. The latter
provides us with unprecedented information about the quantitative
emergence of supersymmetry in this class of models.

\begin{figure}[t] 
\centering
  \includegraphics[width=0.8\textwidth]{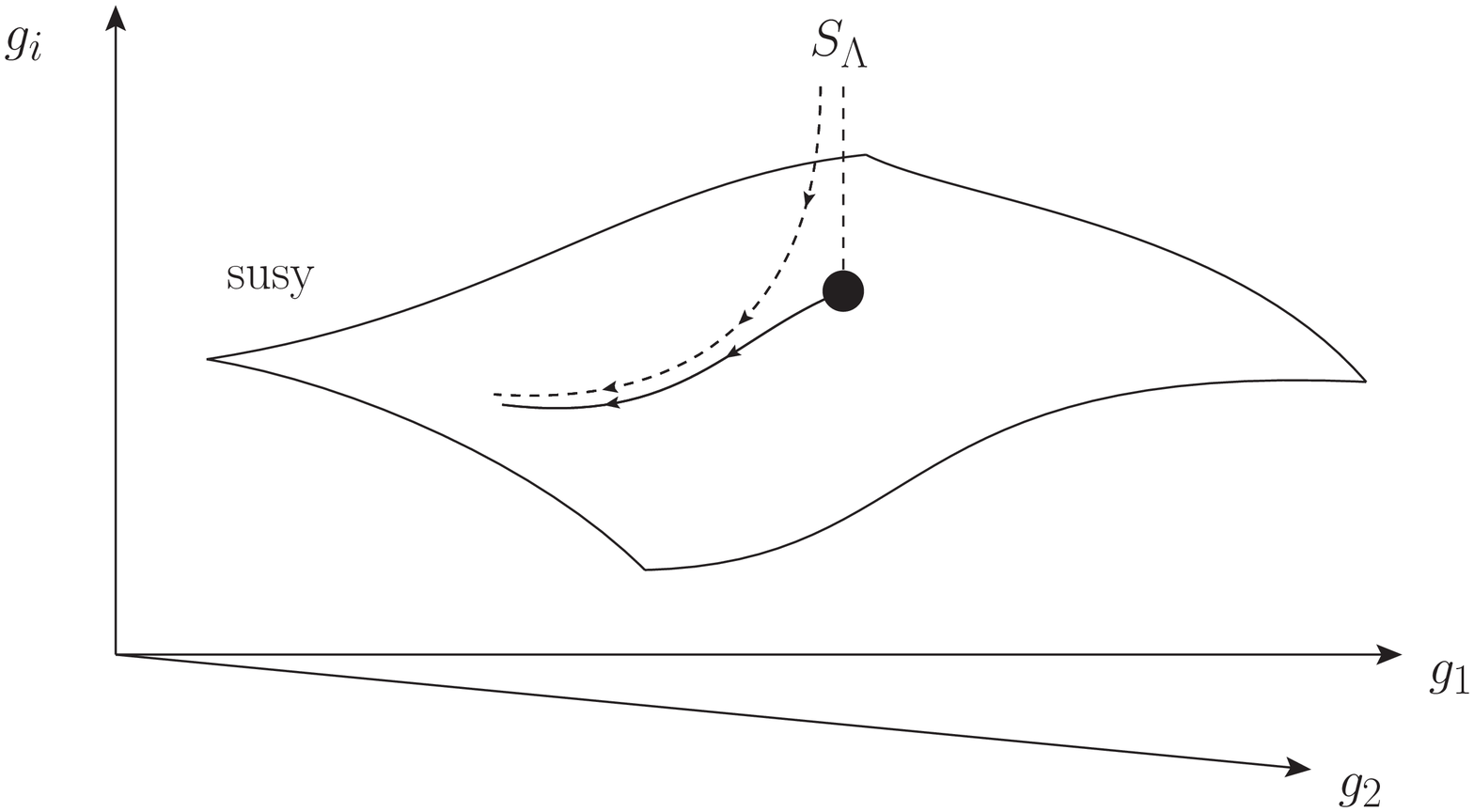}
  \caption{ Illustration of the emergence of supersymmetry in an
    abstract theory space parametrized by couplings
    $\left\{g_i\right\}$. The black dot depicts an RG fixed point
    which lies in the hypersurface of supersymmetric coupling values.
    A manifestly invariant formulation of the theory results in an RG
    flow within the supersymmetric hypersurface, yielding the solid line
    with arrows pointing towards the IR. If all perturbations
    orthogonal to the symmetric hypersurface are RG irrelevant, then the RG
    flow of any nonsymmetric model with microscopic action $S_\Lambda$
    is attracted by the symmetric hypersurface. Thus, supersymmetry
    emerges as a long-range phenomenon.}
\label{fig:IntroSketch}
\end{figure}

It is instructive to abstract the general picture within a heuristic
notion of theory space: consider Lagrangian field theories on the
level of space spanned by all possible operators arising from a given
field content and coordinates given by corresponding coupling values
$g_i$, cf.\ Fig.~\ref{fig:IntroSketch}. The RG flow from microscopic to
macroscopic physics now corresponds to trajectories in this space,
being initiated in terms of a bare action $S_\Lambda$ at a UV cutoff
$\Lambda$. The physics at a lower momentum scale $k$ is then described
by a Wilsonian effective action which is abstractly denoted by
points along the trajectory parametrized by $k$.

Now, any subset of theories in this space with a higher degree of
symmetry forms an invariant subspace, i.e., a hypersurface in theory
space. If the theory is formulated in a manifestly invariant manner,
the corresponding RG trajectory flows completely inside this
hypersurface, see solid trajectory in Fig.~\ref{fig:IntroSketch}. A
criterion for emergent symmetry can now be formulated as follows: if
RG trajectories that are initiated outside the invariant hypersurface
are attracted by the hypersurface towards infrared (IR) scales, the
long-range physics will be dominated by the symmetries defining the
hypersurface. Symmetry violating processes may still be observable, but
are parametrically suppressed by the degree of IR
attractivity of the invariant hypersurface.

This picture becomes fully quantitative in the presence of fixed
points of the RG. In fact, fermionic or Yukawa models in
(2+1)-dimensions are prototypic in that respect since they have
the potential to feature interacting fixed points similar to but
quantitatively different from the Wilson-Fisher fixed point in scalar
theories. Fixed points can be classified according to their number and
nature of relevant (IR-repulsive) directions and corresponding critical
exponents. If the model of higher symmetry, say supersymmetry,
exhibits a fixed point with relevant directions purely inside the
symmetric hypersurface, any nonsymmetric perturbation outside of it
is attracted by the invariant hypersurface towards the IR, cf.\ dashed
trajectory in Fig.~\ref{fig:IntroSketch}. Hence, even nonsymmetric
microscopic models feature symmetric long-range observables: the
symmetry is emergent as a result of the fluctuation-driven
RG flow.\footnote{We emphasize that these requirements ensure that the symmetry emerges \emph{naturally} in the long-range regime.
If the critical point is symmetric, but there are nonsymmetric relevant deformations,
it is still possible to observe the symmetry in the infrared,
provided that additional parameters are tuned to criticality.
This can be illustrated using the model considered in this work:
if a parity symmetry is not imposed,
then supersymmetry can be observed only after tuning
the mass of the fermionic fluctuations to zero.
}

To confirm emergent symmetry as a universal property of a certain
model class, two properties have to be verified: first, the model
should have at least one nontrivial fixed point inside the invariant
hypersurface. Second, all RG relevant directions have to be inside the
symmetric hypersurface as well, whereas perturbations outside the
hypersurface must be irrelevant and thus die out towards the IR. The
second property makes clear that it is not sufficient to study the
manifestly symmetric theory, but the embedding into the non-symmetric
theory space is fundamental to verify and quantify emergence of the
symmetry.

The paper is organized as follows: In Sect.~\ref{section-intro-actions} we introduce the effective action of the Yukawa model under consideration,
investigate the nature of its parity symmetry, and show how to reformulate it
as an explicitly broken supersymmetric model by including an auxiliary field which completes the Yukawa fields into a supermultiplet.
In Sect.~\ref{section-methods} we summarize our two main strategies to study the scaling behavior of the Yukawa model in the light of the results of the preceding section,
and give some further direction to navigate this paper.
Sect.~\ref{section-general-rg-properties} discusses the general form of an RG step, both with and without the auxiliary field.
Sections~\ref{susy-model-section} and~\ref{GNY-section} study the RG flow of the ${\cal N}=1$ model and the Yukawa model respectively,
while Sect.~\ref{section-relations} shows how the former is embedded in the latter at criticality.
Sect.~\ref{section-susy-breaking-flow} gives the RG flow of the broken supersymmetric model and completes the mapping initiated in the previous section.
In Sect.~\ref{section-conclusions} we give the numerical estimates for the critical exponents and compare them with the literature,
while also providing some conclusion and future prospects for our work.

\section{Effective actions, local potentials and symmetries}\label{section-intro-actions}

We work with the language of Lagrangian field theory, using effective actions for the parametrization 
of the theory and the discussion of symmetries. The effective action used in the following can be 
thought of in the spirit of Landau-Ginzburg-Wilson actions, obtained from a suitable coarse graining. 
Within the non-perturbative functional RG used and described in the appendices, a precise connection 
exists to the full 1PI effective action \cite{Wetterich:1992yh,Codello:2013bra,Codello:2017hhh}.

Let us consider a truncation of the effective action of a
general Yukawa model of the form
\begin{equation}\label{truncation_yukawa}
 \begin{split}
  S^{\rm Y}[\varphi,\psi,\bar{\psi}] &= \int {\rm d}^3x \Bigl\{
  \frac{1}{2}\partial_\mu\varphi\partial^\mu\varphi + \frac{i}{2}\bar{\psi}\slashed\partial\psi + 
  U(\varphi)+ \frac{1}{2} H(\varphi) \bar{\psi}\psi
  \Bigr\},
 \end{split}
\end{equation}
in which we introduced a local effective potential $U(\varphi)$ for a real scalar field $\varphi$,
and a function $H(\varphi)$ mediating a Yukawa-type interaction between $\varphi$ and a 
Majorana spinor $\psi$. Even though we mostly work in Euclidean spacetime,
the relevant symmetries are those of the Minkowskian version of the model. For the latter, the action is invariant 
under a parity transformation $x\to\tilde x$, by which $x_2$ changes sign
and the fields transform as
\begin{equation}\label{discrete_chiral}
  \varphi(x) \to -\varphi(\tilde x)\,,
  \quad \psi(x) \to i\gamma^2\psi(\tilde x)\,,
  \quad \bar{\psi} \to -i\bar{\psi}(\tilde x)\gamma^2,
\end{equation}
provided we constrain $U(\varphi)$ and $H(\varphi)$ to be even and odd functions respectively.
The scalar field $\varphi$ can be understood as order parameter for the 
discrete $\Z_2$ symmetry. Note that the parity transformation \eqref{discrete_chiral}
maps Majorana fermions into Majorana fermions.

We find it convenient to introduce a new parametrization of the local functions $U(\varphi)$ and $H(\varphi)$.
Let us first separate the zero point energy from the effective potential $U(\varphi) = U_0 + V(\varphi)$
with the condition that $V(\varphi)$ is zero at the minimum $\varphi_0$ of the potential.
For an effective potential that is bounded from below we have $V(\varphi)\geq 0$ and $V(\varphi_0)=0$.
The zero point energy $U_0$ simply amounts to a global shift in the energy spectrum and will be ignored in the following discussion.
We then introduce two new local functions $W(\varphi)$ and $Y(\varphi)$, which are defined implicitly as
\begin{equation}\label{potentials_redefinition}
 \begin{split}
  V(\varphi) = \frac{1}{2}W'(\varphi)^2\,, &\qquad
  H(\varphi) = W''(\varphi)+Y(\varphi).
 \end{split}
\end{equation}
By construction $W(\varphi)$ is a real valued function because $V(\varphi)\geq 0$.
In the new parametrization the action of the Yukawa model becomes
\begin{equation}\label{truncation_yukawa_susy_on}
 \begin{split}
  S^{\rm Y}[\varphi,\psi,\bar{\psi}] &= \int {\rm d}^3x \,\Bigl\{
  \frac{1}{2}\partial_\mu\varphi\partial^\mu\varphi + \frac{i}{2}\bar{\psi}\slashed\partial\psi + 
  \frac{1}{2}W'(\varphi)^2+\frac{1}{2}W''(\varphi)\bar{\psi}\psi + \frac{1}{2}Y(\varphi) \bar{\psi}\psi
  \Bigr\}\,.
 \end{split}
\end{equation}
The new functions $W(\varphi)$ and $Y(\varphi)$ can be constrained by parity
\eqref{discrete_chiral} to be odd functions of the field.

Let us introduce an auxiliary field $F$ whose purpose is to complete the fields $(\varphi,\psi)$ into a 
full supermultiplet $\Phi=(\varphi,\psi,F)$. Using the path integral
\begin{equation}\label{F_path_integral}
 \begin{split}
  {\rm e}^{i S^{\rm Y}[\varphi,\psi,\bar{\psi}]} &= \int \! DF \, {\rm e}^{iS[\varphi,\psi,\bar{\psi},F]},
 \end{split}
\end{equation}
we define a new effective action as
\begin{equation}\label{truncation_yukawa_susy_off}
 \begin{split}
  S[\varphi,\psi,\bar{\psi},F] &=
  \int {\rm d}^3x \,\Bigl\{ \frac{1}{2}\partial_\mu\varphi\partial^\mu\varphi + \frac{i}{2}\bar{\psi}\slashed\partial\psi
  -\frac{1}{2}F^2+F W'(\varphi)+\frac{1}{2}W''(\varphi)\bar{\psi}\psi +\frac{1}{2}Y(\varphi) \bar{\psi}\psi\Bigr\}\,.
 \end{split}
\end{equation}
The new field $F$ is a scalar field which we require to be invariant under the transformation \eqref{discrete_chiral}
in order to naturally extend any parity property of the actions \eqref{truncation_yukawa} and \eqref{truncation_yukawa_susy_on}.
At any moment, one can make use of the equations of motions $F=W'(\varphi)$ in the effective 
action \eqref{truncation_yukawa_susy_off} to eliminate the auxiliary field $F$ and return to the formulation 
\eqref{truncation_yukawa_susy_on}.

The new effective action can be understood as the sum of an effective action which is manifestly ${\cal N}=1$ supersymmetric
\begin{equation}\label{truncation_susy_off}
 \begin{split}
  S^{{\cal N}=1}[\varphi,\psi,\bar{\psi},F] &=
  \int {\rm d}^3x \Bigl\{ \frac{1}{2}\partial_\mu\varphi\partial^\mu\varphi 
  + \frac{i}{2}\bar{\psi}\slashed\partial\psi -\frac{1}{2}F^2+F W'(\varphi)+\frac{1}{2}W''(\varphi)\bar{\psi}\psi \Bigr\}\,,\\
 \end{split}
\end{equation}
and a term that breaks supersymmetry explicitly
\begin{equation}\label{truncation_breaking_term}
 \begin{split}
  S^{\rm BR}[\varphi,\psi,\bar{\psi}] &=
  \frac{1}{2}\int {\rm d}^3x\, Y(\varphi) \bar{\psi}\psi\,.
 \end{split}
\end{equation}
The off-shell ${\cal N}=1$ supersymmetry transformations
\begin{equation}\label{susy_transformations_on_shell}
 \delta \varphi = \bar{\theta}\psi\,, \qquad \delta\psi = \left(F+i\slashed\partial\varphi\right)\theta\,, \qquad 
 \delta F = i\bar{\theta}\slashed\partial \psi\,,
\end{equation}
contain an infinitesimal Majorana spinor $\theta$ as parameter.
Analogously to the expectation value of $\varphi$ as an order parameter for  parity, the expectation value of the auxiliary field $F$ is  
an order parameter of ${\cal N}=1$ supersymmetry.

It is important to notice that ${\cal N}=1$ supersymmetry is not a purely mathematical notion that was introduced 
by hand with the field $F$. To see this let us integrate out the field $F$ with the path integral \eqref{F_path_integral}.
Using the equation of motion $F=W'(\varphi)$ in \eqref{susy_transformations_on_shell} we obtain the 
on-shell ${\cal N}=1$ supersymmetric transformations
\begin{equation}\label{susy_transformations_off_shell}
 \delta \varphi = \bar{\theta}\psi\,, \qquad \delta\psi = \left(W'(\varphi)+i\slashed\partial\varphi\right)\theta\,.
\end{equation}
The first four terms of \eqref{truncation_yukawa_susy_on} are invariant 
under this symmetry, while the last term, which corresponds to \eqref{truncation_breaking_term}, breaks it explicitly.
This is of course in complete analogy with the invariance property exhibited by \eqref{truncation_susy_off}.

Before concluding this Section it is important to discuss the mechanisms of symmetry breaking that are 
induced by the action \eqref{truncation_breaking_term}.
There are in principle two different symmetry breaking mechanisms induced by $Y(\varphi)$:
On the one hand, if $Y(\varphi)$ is not an odd function of $\varphi$ then both parity and ${\cal N}=1$ 
supersymmetry are broken explicitly by \eqref{truncation_breaking_term}. This is the case, for example, in which $Y(\varphi)$ is 
constant and we obtain a correction to the mass term for the Majorana fermions $Y(\varphi)\sim \delta m_\psi $,
which is not balanced by a corresponding change in the mass of $\varphi$.
The correction to the mass is expected to provide a relevant deformation to the theory's spectrum, and thus it is a 
feature of the theory that is expected to become increasingly important in the infrared (at low energies 
or large scales, as compared to the energy scale provided by  $\delta m_\psi $). On the other hand, if $Y(\varphi)$ is an 
odd function of $\varphi$, parity can be preserved while only ${\cal N}=1$ supersymmetry is broken.
This can be achieved if $Y(\varphi)$ entails an additional Yukawa interaction $Y(\varphi)\sim y\varphi$ 
on top of the supersymmetric one produced by $W(\varphi)$. We will see in the following 
that this new Yukawa interaction contributes with a deformation of the theory's spectrum which becomes 
irrelevant because of statistical fluctuations. Differently from the mass term, an irrelevant breaking 
term \eqref{truncation_breaking_term} is expected to be increasingly less important in the infrared.

We can draw an interesting conclusion based on the last paragraph:
\emph{If} a system such as \eqref{truncation_yukawa} contains \emph{massless} Majorana excitations which are
coupled to some scalar order parameter with a Yukawa-type interaction,
\emph{then} the resulting system is expected to display supersymmetry as an emergent feature in the infrared.
The precise implications of the emergence of supersymmetry will be clarified in the following.

\section{Perturbative vs non-perturbative RG}\label{section-methods}

We investigate the scale dependence of all the systems \eqref{truncation_yukawa_susy_on}, \eqref{truncation_yukawa_susy_off} 
and \eqref{truncation_susy_off} by means of two different and rather complementary RG methods,
one based on perturbation theory \cite{ODwyer:2007brp} which was dubbed functional perturbative RG in \cite{Codello:2017hhh},
and another based on a non-perturbative RG equation that goes under the name of Wetterich 
equation and is known as functional RG \cite{Wetterich:1992yh}. We illustrate the most important results of the paper by means of 
functional perturbation theory from  Sect.~\ref{section-general-rg-properties} to Sect.~\ref{section-susy-breaking-flow},
because it makes our discussion much more transparent. In the conclusions 
of Sect.~\ref{section-conclusions}, we also give all numerical estimates obtained with the non-perturbative method,
and summarize the main non-perturbative formulas in Appendix~\ref{susy-breaking-appendix} and 
Appendix~\ref{GNY-appendix}. All perturbative results given in the main text of the paper can be derived from
a standard application of perturbation theory and dimensional regularization, but are also
fully contained as the logarithmic terms of the non-perturbative results given in the appendices.
Both procedures are described in more detail in \cite{Codello:2017hhh}.

The renormalization of the Yukawa system \eqref{truncation_yukawa_susy_on} can be performed using standard 
dimensional regularization and perturbation theory. For this purpose,
we analytically continue the action to $d$ dimensions and identify the quartic self-interaction 
of $V(\varphi)$ and the Yukawa interaction of $H(\varphi)$ as the critical couplings of the model.
The upper critical dimension is $d_c=4$, so the theory's critical properties can be computed as a 
systematic expansion in the parameter $\epsilon$, which represents the displacement from the upper critical 
dimension $d=4-\epsilon$. The three dimensional system corresponds to $\epsilon=1$ which is arguably 
outside a reliable range of perturbation theory, but it may still give reasonable estimates for the critical exponents.
The leading order of the $\epsilon$ expansion of \eqref{truncation_yukawa_susy_on} is given in 
Section \ref{GNY-section} and agrees with \cite{Rosenstein:1993zf,Fei:2016sgs}. 

The Yukawa system can also be studied with functional RG methods directly in three dimensions.
The analysis of the RG with this method has been carried over extensively in \cite{Vacca:2015nta} and 
will not be reproduced here in its entirety. However, we used the results of \cite{Vacca:2015nta} 
and \cite{Zanusso:2009bs,Vacca:2010mj,Zanusso:PhD} to reproduce the perturbative results presented 
in \ref{GNY-section}, while summarizing the non-perturbative RG beta functions in Appendix~\ref{GNY-appendix}.
For this comparison it is important to realize that the analytic continuation of \emph{one} Majorana fermion 
in three dimensions corresponds to $N_f=1/4$ Dirac spinors in four dimensions, and therefore the 
representation of the Clifford algebra of \eqref{truncation_yukawa} must be modified to take this fact into account.

The perturbative renormalization of the ${\cal N}=1$ supersymmetric model \eqref{truncation_susy_off} 
is presented in Sect.~\ref{susy-model-section} and can be derived either from the results of \cite{Fei:2016sgs} 
or from the perturbative part of the results of \cite{Hellwig:2015woa}. The perturbative effects of the 
supersymmetry breaking term \eqref{truncation_breaking_term} are presented in Section \ref{section-susy-breaking-flow}.
The non-perturbative RG flow of \eqref{truncation_susy_off} and \eqref{truncation_yukawa_susy_off} is 
summarized in Appendix \ref{susy-breaking-appendix}.

\section{Properties of the RG flow and $F$-field redefinition}\label{section-general-rg-properties}

In this Section, we want to outline some general properties of the RG flow of
the system described by the effective action \eqref{truncation_yukawa_susy_off}
containing the breaking term.
These properties do not depend on the specific RG method, and will prove crucial in what follows.
Let us begin by introducing an RG scale $k$ and considering the RG transformation 
of $S^{{\cal N}=1}[\varphi,\psi,\bar{\psi},F]$ for an infinitesimal change $k\to k-\delta k$
\begin{equation}\label{rg-step-susy}
 \begin{split}
  S^{{\cal N}=1} & \to S^{{\cal N}=1} - 
  \frac{\delta k}{k} \int {\rm d}^dx \Bigl\{
  \beta_W' F+\frac{1}{2}\beta_W''\bar{\psi}\psi \Bigr\}\,,
 \end{split}
\end{equation}
where we have introduced the beta function 
$\beta_W \equiv k\partial_k W(\varphi)$
of the superpotential $W(\varphi)$.
Here we implicitly assume that a supersymmetric action flows into another supersymmetric action,
a fact which is confirmed by direct observation when using $\overline{\rm MS}$ methods \cite{Fei:2016sgs},
but that can be proven explicitly using functional RG methods and choosing 
a supersymmetric cutoff \cite{Synatschke:2008pv,Synatschke:2009nm,Hellwig:2015woa,Mastaler:2012xg,Heilmann:2014iga}.

The situation complicates slightly when studying an RG transformation of $ S[\varphi,\psi,\bar{\psi},F]$.
In general such a flow is parametrized by three functions, which we choose as
\begin{equation}\label{rg-step-broken}
 \begin{split}
  S & \to S - 
  \frac{\delta k}{k} \int {\rm d}^dx \Bigl\{
  A(\varphi) F +\frac{1}{2} B(\varphi)\bar{\psi}\psi +C(\varphi) + \beta_{U_0}
  \Bigr\}\,.
 \end{split}
\end{equation}
In \eqref{rg-step-broken}, we have already separated the flow of the zero point energy $\beta_{U_0}$, which can be neglected for the following discussion.
The function $C(\varphi)$ is thus normalized to be zero at the ground state $\varphi_0$ (see also the discussion of Sect.~\ref{section-intro-actions}).

\begin{figure}[t]
\centering
  \includegraphics[width=0.65\textwidth]{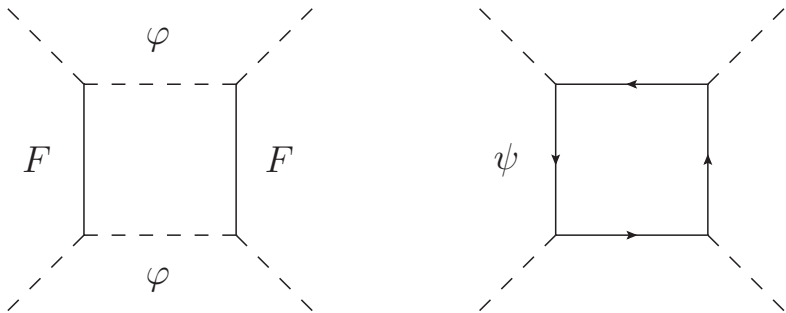}
  \caption{
  Diagrams responsible for the generation of $C(\varphi)\sim \varphi^4$ of \eqref{rg-step-broken} in perturbation theory.
  The diagram on the right has an additional minus sign due to the fermion loop, therefore the two diagrams tend 
  to balance each other $C(\varphi)\propto W''(\varphi)^4-H(\varphi)^4$. In fact, if the Yukawa function $H(\varphi)$ 
  satisfies the requirement of supersymmetry $H(\varphi)=W''(\varphi)$, then the two diagrams sum to zero
  and the RG step reduces to \eqref{rg-step-susy}.}
\label{diagrams}
\end{figure}

The supersymmetry breaking term \eqref{truncation_breaking_term} has two notable effects:
On the one hand the functions $A(\varphi)$ and $B(\varphi)$ are not related by a simple differentiation 
as in  \eqref{rg-step-susy}.
On the other hand the new function
$C(\varphi)$ arises as 
a scalar self-interaction which  is not mediated by  neither $F$ nor $\bar{\psi}\psi$ (compare \eqref{rg-step-broken} with 
\eqref{truncation_yukawa_susy_off}).
The monomial $C(\varphi)$
poses quite some problem if one attempts an interpretation of the right hand side of \eqref{rg-step-broken} 
in terms of beta functions, because it is 
not clear to what function's flow should $C(\varphi)$ be attributed to
from the point of view of the off-shell supersymmetric action \eqref{truncation_yukawa_susy_off}.
This problem occurs for both perturbative \emph{and} non-perturbative methods.
In fact, close to the upper critical dimension the function $C(\varphi)$ plays the role
of an additional $\varphi^4$ interaction \emph{besides} the one already present in the superpotential:
The one loop diagrams whose renormalization generates $C(\varphi)$ in \eqref{rg-step-broken} 
are shown in Fig.\ \ref{diagrams}.

The physical reason for this problem is the following: while the
field $F$ is defined at the scale $k$ as an auxiliary field for
exactly parametrizing certain boson and fermion operators, it does
no longer do so at the scale $k-\delta k$. Fluctuations have
modified the corresponding boson and fermion operators, such that
the auxiliary field $F$ has to be adjusted accordingly. Within the
functional RG, this problem can be solved by scale-dependent field
transformations
\cite{Gies:2001nw,Pawlowski:2005xe,Gies:2006wv,Floerchinger:2009uf}.
In the present case, we can, in fact, eliminate $C(\varphi)$ by
an appropriate redefinition of the $F$ field along the RG.  By
demanding that the structure of the path-integral
\eqref{F_path_integral} holds at \emph{any} scale $k$ instead of a
given ultraviolet scale $\Lambda \gg k$, we can infer that an
appropriate scale-dependent nonlinear redefinition $\beta_F \equiv
k\partial_k F = D(\varphi)$ of $F$ should be coupled to the flow
\begin{equation}\label{rg-step-broken2}
 \begin{split}
  S & \to S - 
  \frac{\delta k}{k} \int {\rm d}^dx \Bigl\{
  A(\varphi) F +\frac{1}{2} B(\varphi)\bar{\psi}\psi +C(\varphi)
  - \frac{\delta S}{\delta F} D(\varphi) \Bigr\}\,,
 \end{split}
\end{equation}
in order to cancel the contribution coming from $C(\varphi)$.
Using the explicit form of \eqref{truncation_yukawa_susy_off} in \eqref{rg-step-broken2},
we can see that the choice
\begin{equation}\label{F-field-flow}
 \begin{split}
  \beta_F  &= D(\varphi) = \frac{C(\varphi)}{W'(\varphi)}
 \end{split}
\end{equation}
returns an RG step which agrees with the structure of \eqref{truncation_yukawa_susy_off} itself
\begin{equation}\label{rg-step-broken3}
 \begin{split}
  S & \to S - 
  \frac{\delta k}{k} \int {\rm d}^dx \Bigl\{
  \Bigl(A(\varphi) -\frac{C(\varphi)}{W'(\varphi)}\Bigr)F +\frac{1}{2} B(\varphi)\bar{\psi}\psi \Bigr\}\,,
 \end{split}
\end{equation}
and which can be used to attribute beta functions to the superpotential $W(\varphi)$ and 
to $Y(\varphi)$ as they appear in \eqref{truncation_yukawa_susy_off}.

The nonlinearity of \eqref{F-field-flow} is very important in protecting the auxiliary role 
of the field $F$. To see this, let us first recall the equation of motion of $F$
\begin{equation}\label{F-eq-of-motion}
 \begin{split}
 F=W'(\varphi)\,,
 \end{split}
\end{equation}
and consider it at a minimum $\varphi_0$ of the on-shell potential $V(\varphi)=W'(\varphi)^2/2$ . 
Since $V(\varphi)$ vanishes at its minimum $\varphi_0$,
we have $W'(\varphi_0)=0$. Therefore, if we expand around the 
minimum $\varphi_0$, then the only way to have an expansion in the field $F$ which is consistent 
with \ref{F-eq-of-motion} is to choose $F_0=0$ as the expansion point. Using L'H\^{o}pital's rule 
on \eqref{F-field-flow},  it is easy to check that
\begin{equation}\label{F-field-hopital}
 \begin{split}
  0  &= \frac{C'(\varphi_0)}{W''(\varphi_0)}
 \end{split}
\end{equation}
ensures that the condition $F=0$ is preserved along the flow at the minimum. This will be true for both the 
perturbative and the non-perturbative redefinitions, which are given in Sect.~\ref{section-susy-breaking-flow} and 
Appendix~\ref{susy-breaking-appendix} respectively.

\section{RG flow of the ${\cal N}=1$ model}
\label{susy-model-section}

Let us now turn to the renormalization of the manifestly off-shell supersymmetric 
action \eqref{truncation_susy_off}. It is expected that there exists a critical point for any dimension $2\leq d <4 $, 
with four being the upper critical dimension \cite{Fei:2016sgs}. The critical point can also be understood as 
the ${\cal N}=1$ supersymmetric generalization of the well-known Wilson-Fisher fixed point. The non-perturbative 
renormalization group flow of the superpotential $W(\varphi)$ has been studied in 
\cite{Mastaler:2012xg,Heilmann:2014iga} for the three dimensional case, and it has been generalized 
in \cite{Hellwig:2015woa} to the whole range $2\leq d <4 $. The latter work contains
leading order of the universal contributions to the flow close to the upper critical dimension $d=4-\epsilon$,
which agrees with the RG flow presented in \cite{Fei:2016sgs} and which we shall use in our work.

We introduce the dimensionless superpotential by measuring its dimensionful counterpart in units of the RG scale $k$
\begin{equation}\label{dimensionless-superpotential}
\begin{split}
 w(\varphi) &= k^{-d+1} W(\varphi \, k^{d/2-1} Z^{-1/2})\,, 
\end{split}
\end{equation}
where $Z$ is a wave-function normalization for the superfield whose RG flow yields the anomalous dimension 
$\eta=- k \partial_k \log Z$. It should be understood that now $\varphi$ denotes the dimensionless field, 
whose scaling is corrected by $\eta$, and that the anomalous dimension can be fixed by the requirement 
that the fields of the supermultiplet of \eqref{truncation_susy_off} have canonically normalized propagators.
The perturbative beta function of $w(\varphi)$ is
\begin{equation}\label{susy-flow}
 \begin{split}
  \beta_w  &= -(d-1)w(\varphi)+ \frac{d-2+\eta}{2}\varphi w'(\varphi)+\frac{1}{3(4\pi)^2}w''(\varphi)^3\\
  &= -3 w(\varphi) +\varphi w'(\varphi) + \epsilon \Bigl( w(\varphi)-\frac{1}{2}\varphi w'(\varphi)\Bigr)+ \frac{\eta}{2}\varphi w'(\varphi) +\frac{1}{3(4\pi)^2}w''(\varphi)^3 \,.\\
 \end{split}
\end{equation}
In the first line we distinguish the first three terms, that correspond to the scaling induced by using $k$ as unit, 
from the last two, that express the effects of quantum fluctuations of $\varphi$ and of the renormalization of $W(\varphi)$.
In the second line we further rewrote the result by using $d=4-\epsilon$, so that the scaling terms are separated 
into the standard canonical scaling of $4-\epsilon$ dimensions and the quantum induced effects of having an 
anomalous dimension $\eta$. The anomalous dimension is related to the superpotential by
\begin{equation}\label{susy-eta}
 \begin{split}
  \eta & = \frac{1}{(4\pi)^2}w'''(0)^2\,.
 \end{split}
\end{equation}
The third derivative of the superpotential is here evaluated at the
minimum of the effective potential: We confirm below that the
critical potential has always $\varphi=0$ as minimum within
perturbation theory, demonstrating self-consistency of this
formula. In functional RG computations, instead, the minimum might
shift to a value $\varphi_0 \neq 0$ and it is conventional to use this minimum when evaluating $\eta$.
We refer the Reader to appendix \ref{appendix-VS} for a more detailed explanation on
how to derive functional perturbative RG flows and their relation with non-perturbative flows.

The RG flow \eqref{susy-flow} can be understood as the generating functional of the beta functions
of the couplings to the local operators that are obtained from raising the field $\varphi$ to an arbitrary power \cite{Codello:2017hhh}.
To clarify the relation between \eqref{susy-flow} and the beta function presented in \cite{Fei:2016sgs},
let us introduce the critical coupling $\lambda$, which is canonically dimensionless in four dimensions,
and insert the following parametrization for the
critical potential $w(\varphi)= \frac{\lambda}{3!}\varphi^3$ in \eqref{susy-flow}. The coupling $\lambda$ governs the perturbative 
expansion  of the system and is canonically dimensionless at the  upper critical dimension.
The flow of the superpotential can then be used to determine the beta function of $\lambda$ as $\beta_w= 
\frac{\beta_\lambda}{3!}\varphi^3$. We end up with
\begin{equation}\label{beta-lambda}
 \begin{split}
  \beta_\lambda  = -\frac{\epsilon}{2} \lambda + \frac{7}{2(4\pi)^2}\lambda^3\,,\qquad
  \eta  = \frac{1}{(4\pi)^2}\lambda^2\,.
 \end{split}
\end{equation}
This simple construction does not yet take into account all dimensionful couplings, which can be included
through a more general parametrization of the potential $w(\varphi) = \frac{\lambda}{3!}\varphi^3 
+ \sum_{n\neq 3} \frac{\lambda_n}{n!} \varphi^n$. 
Since our discussion is based on perturbation theory, all $\lambda_n$ are expected to be zero at criticality, 
thus the flow \eqref{susy-flow} can straightforwardly be related to its critical exponents $\theta_n$ through
\begin{equation}\label{susy-flow-gamma-functions}
 \begin{split}
  \beta_w  &= \frac{1}{3!} \beta_\lambda(\lambda)\varphi^3 - 
  \sum_{n\neq 3} \frac{\lambda_n}{n!} \theta_{n}(\lambda) \varphi^n + {\cal O}(\lambda_n\lambda_{n'})\,.
 \end{split}
\end{equation}
In functional RG applications non-critical couplings might be non-zero at criticality, so the 
above relation is more complicated \cite{Codello:2013bra} and the computation of the critical exponents 
might require the diagonalization of the stability matrix. The function $\beta_\lambda=\beta_\lambda(\lambda)$ 
and the exponents $\theta_n=\theta_n(\lambda)$ depend on the coupling $\lambda$ alone
and control the scaling of the corresponding \emph{dimensionless} operators.

At criticality the system exhibits scale invariance, thus it must realize $\beta_w=0$. This condition can be achieved 
by fine tuning the coupling $\lambda$: The beta function \eqref{beta-lambda} has a nontrivial fixed point for
\begin{equation}\label{susy-fp}
 \begin{split}
 \frac{\lambda^{\star 2}}{(4\pi)^2}=\frac{\epsilon}{7}\,.
 \end{split}
\end{equation}
The fixed point allows us to trade the $\lambda$ dependence of the exponents for an epsilon expansion.
We obtain the spectrum
\begin{equation}\label{susy-spectrum}
 \begin{split}
  \theta_n
  &= 3-n -\epsilon - \frac{1}{7}n(n-4)\epsilon-3\,\eta\, \delta_{n,3} \qquad {\rm for}\quad n\geq 0\\
  &= \Bigl\{ 3-\epsilon\,, 2-\frac{4}{7}\epsilon\,, 1-\frac{3}{7}\epsilon\,,-\epsilon\,,\dots\Bigr\}\,.\\
 \end{split}
\end{equation}
The anomalous dimension is
\begin{equation}\label{susy-eta2}
 \begin{split}
  \eta &=\frac{\epsilon}{7}\,.
 \end{split}
\end{equation}
We arranged the spectrum such that the special case $n=3$ includes the critical exponent associated with 
the $\varphi^3$ deformation. Its negative is sometimes denoted $\omega = -\theta_3 = \beta_\lambda'(\lambda^\star)
=\epsilon$ and its value differs from what one would naively obtain from setting $n=3$ in the first terms 
of the first line of \eqref{susy-spectrum} because of the presence of the anomalous dimension in the beta  function \eqref{beta-lambda}.
The ellipsis denotes subsequent irrelevant operators for $n>3$ which are subject to further mixing with derivative operators \cite{Codello:2017hhh}.
Therefore, our formula is not expected to give the correct result for their leading expansion in $\epsilon$.
The inclusion of such derivative operators in the analysis is, of course, possible both perturbatively and non-perturbatively \cite{ODwyer:2007brp}.

Let us carefully spell out the physical meaning of the exponents.
It is possible to prove in general that $\delta w\sim \varphi$ and $\delta w\sim w'(\varphi)\sim\varphi^2$ are allowed 
deformations of \eqref{susy-flow} with critical exponents $\theta_1=(d-\eta)/2$ and $\theta_2=\Delta=(d-2+\eta)/2$ 
respectively (see for example the Appendix of \cite{Hellwig:2015woa}). The critical exponent $\theta_2$ is the 
quantum-corrected scaling dimension of the field operator $\varphi$, which is sometimes denoted $\Delta$,
and is related to the anomalous dimension by the scaling relation $\theta_2=(d-2+\eta)/2$.
The critical exponent $\theta_1$ of \eqref{susy-spectrum} corresponds to a linear deformation of $w(\varphi)$.
and satisfies another scaling relation $\theta_1=(d-\eta)/2$.
As the leading nontrivial relevant exponent, $\theta_1$ governs the approach to criticality
and thus the onset of order parameters across the corresponding phase transition.
It plays the same role for the superpotential as does the (inverse) correlation length exponent $\nu^{-1}$
for the on-shell potential. Hence, let us write $\nu_W^{-1}:=\theta_1$ \cite{Gies:2009az}.
Finally, the exponent $\theta_0$ trivially corresponds to the overall scaling $k^{-d+1}$ of the superpotential,
which is never modified by quantum fluctuations and bears no physical meaning. 

Together these relations prove that the spectrum of the relevant operators of the ${\cal N}=1$ model
can be completely determined by the knowledge of a single exponent, which we take to be the anomalous dimension $\eta$.
In particular, the scaling of the deformation of the superpotential
is related to $\eta$ by the so-called superscaling relation \cite{Gies:2009az}
\begin{equation}\label{superscaling-relation}
 \begin{split}
  \nu_W^{-1}=\frac{d-\eta}{2}\,.
 \end{split}
\end{equation}
This relation is the most important effect of supersymmetry, and marks a big difference to what happens in the Wilson-Fisher universality class.
For the latter, independent knowledge about the two exponents $\eta$ and $\nu$ is needed to fully characterize the relevant properties of the spectrum.
The ability to determine all relevant critical exponents from $\eta$ alone
is the smoking gun of the emergence of supersymmetry.
The following analysis of the Yukawa model ultimately will lead us to the conclusion
that the exponent $\nu_W$ can be identified with the correlation length exponent, $\nu\equiv\nu_W$.

While Eq.~\eqref{superscaling-relation} holds perturbatively for obvious reasons, it is nontrivial to satisfy the scaling relation with nonperturbative methods
in a truncated theory space and away from its upper critical dimension.
An important feature of the numerical estimates obtained with non-perturbative RG methods and reported in Sect.~\ref{section-conclusions}
is that Eq.~\eqref{superscaling-relation} can be satisfied either exactly or with an arbitrary numerical accuracy.
In fact, it is easy to prove Eq.~\eqref{superscaling-relation} within a local potential approximation (LPA)
using a simple RG stability analysis and keeping $\eta$ fixed as an external parameter (see for example the appendices of \cite{Hellwig:2015woa,Codello:2017hhh}).
The idea of keeping the anomalous dimension fixed when studying the RG stability has been discussed also in \cite{Boettcher:2015pja} for multicritical systems.
If instead the anomalous dimension is not kept fixed, we expect that \eqref{superscaling-relation}
can be satisfied with arbitrary accuracy for increasing size of the truncation in the space of all possible operators,
as has been observed in \cite{An:2016lni,Zambelli:2016cbw} for the Lee-Yang model.
The accuracy with which the superscaling relation is satisfied thus becomes a benchmark test for all our numerical estimates.

Before concluding this Section, we find convenient to give the explicit form of the deformations of the on-shell effective action
corresponding to the critical exponents $\theta_2$ and $\theta_3$.
They are
\begin{equation}\label{susy-explicit-deformations}
 \begin{split}
  \delta S \propto	\begin{cases}                        
			  \int_x \left(\bar{\psi}\psi+\lambda^\star\varphi^{3}\right) 			&{\rm with \quad}\theta_2 = 1-\frac{3}{7}\epsilon \,,\\
			  \int_x \left(\varphi\bar{\psi}\psi+\frac{\lambda^\star}{2}\varphi^{4}\right)	&{\rm with \quad}\theta_3 = -\epsilon \,.
			\end{cases}
 \end{split}
\end{equation}
As already discussed above, the first deformation breaks the parity \eqref{discrete_chiral} manifestly, while the second is invariant.
Both of them are, of course, invariant under supersymmetry.

\section{RG flow of the Yukawa model} \label{GNY-section}

For the renormalization of \eqref{truncation_yukawa} we will follow a similar procedure as in the previous Section.
Since the self-interaction potential and the Yukawa function are generally independent, we have that \eqref{truncation_yukawa}
can describe a number of critical points, but only one displays non trivial functions for both interactions.
As in the previous Section, the RG flow of $U(\varphi)$ and $H(\varphi)$ can be studied in a multitude of ways.
We conduct our manipulations here using the perturbative flow, which can be either obtained by renormalizing the theory
with critical $\varphi^4$ and Yukawa couplings below four dimensions, or by extracting the universal part of 
the functional RG flow. These two approaches give the same results.

We define the dimensionless potential and Yukawa function
\begin{equation}
\begin{split}
 u(\varphi) &= k^{-d} U(\varphi \, k^{d/2-1}Z^{-1/2}_\varphi)\,, \\ 
 h(\varphi) &= k^{-1} Z_\psi^{-1} H(\varphi \, k^{d/2-1}Z^{-1/2}_\varphi)\,, 
\end{split}
\end{equation}
where $Z_\varphi$ and $Z_\psi$ are separate wave-function normalizations for the fields $\varphi$ and $\psi$,
which are related to the anomalous dimensions $\eta_\varphi=-k\partial_k \log Z_\varphi$ 
and $\eta_\psi=-k\partial_k \log Z_\psi$. Since in the general Yukawa model the fields $\varphi$ and $\psi$ 
are not members of a single supermultiplet, they might have different anomalous dimensions.

The perturbative beta functions of the two functions in $d=4-\epsilon$ are
\begin{equation}\label{GNY-flow}
 \begin{split}
  \beta_u &= -4u + \varphi u' + \epsilon \Bigl( u-\frac{1}{2}\varphi u'\Bigr) 
  + \frac{\eta_\varphi}{2}\varphi u' + \frac{1}{2(4\pi)^2}(u'')^2  - \frac{1}{2(4\pi)^2}h^4\,,\\
  \beta_h &= -h + \varphi h'-\epsilon \frac{1}{2}\varphi h' +\eta_\psi h 
  + \frac{\eta_\varphi}{2}\varphi h' + \frac{2}{(4\pi)^2} h (h')^2\,.
 \end{split}
\end{equation}
A single Majorana field in three dimensions relates to $N_f=1/4$ Dirac spinors in four dimensions \cite{Vacca:2015nta,Fei:2016sgs}.
This has the important consequence that the two anomalous dimensions actually coincide
\begin{equation}\label{GNY-eta}
 \begin{split}
 \eta_\psi=\eta_\varphi = \frac{1}{(4\pi)^2} h'(0)^2\,,
 \end{split}
\end{equation}
and therefore they can in principle be part of the same superfield multiplet.

It is straightforward to repeat the steps of the previous Section that lead to the non-trivial fixed 
point with corresponding spectrum. Using the notation of \cite{Fei:2016sgs}, let us define the critical 
couplings $g_1$ and $g_2$ as
\begin{equation}
 \begin{split}
 h(\varphi) = g_1\varphi\,, \qquad  u(\varphi) = \frac{g_2}{4!}\varphi^4\,.
 \end{split}
\end{equation}
Their beta functions $\beta_{g_1}$ and $\beta_{g_2}$ can be easily extracted from those of the 
dimensionless potentials \eqref{GNY-flow} and agree with the one loop part of \cite{Fei:2016sgs}
\begin{equation}
 \begin{split}
 \beta_{g_1}=-\frac{\epsilon}{2}g_1+\frac{7}{32\pi^2}g_1^3\,,& \qquad \beta_{g_2}=-\epsilon g_2 +\frac{3}{16\pi^2}g_2^2+\frac{1}{8\pi^2}g_2g_1^2-\frac{3}{4\pi^2}g_1^4\,.
 \end{split}
\end{equation}
They have the following fixed point 
\begin{equation}\label{GNY-fp}
 \begin{split}
 \frac{g^{\ast}_1{}^2}{(4\pi)^2}=\frac{\epsilon}{7}\,, \qquad  \frac{g^\ast_2}{(4\pi)^2}=\frac{3\epsilon}{7}\,.
 \end{split}
\end{equation}
The computation of the spectrum is slightly more involved than in the previous Section because of the 
mixing between the operators of $u(\varphi)$ and $h(\varphi)$.
There are two sets of critical exponents, $\theta_{n,1}$ for $n\geq 0$ and $\theta_{n,2}$ for $n \geq 3$.
Exponents corresponding to the same value of $n$ result from the mixing of the operators $\varphi^n$ and $\varphi^{n-3}\bar{\psi}\psi$.
We find
\begin{equation}\label{GNY-spectrum}
 \begin{split}
  \theta_{n,1}
  &=  4-n-\epsilon-\frac{3}{14}n\left(n-3\right)\epsilon \,, \qquad {\rm for}\quad n \geq 0\,,\\
  &=  \Bigl\{ 4-\epsilon\,,3-\frac{4}{7}\epsilon\,, 2-\frac{4}{7}\epsilon\,, 1-\epsilon\,,-\frac{13}{7}\epsilon\,,\dots\Bigr\}\,;\\
  \theta_{n,2}
  &=  4-n-\frac{1}{7}n\epsilon -3\,\eta\, \delta_{n,4}\,, \qquad {\rm for}\quad n \geq 3\,,\\
  &=  \Bigl\{ 1-\frac{3}{7}\epsilon\,,-\epsilon\,,\dots\Bigr\}\,.
 \end{split}
\end{equation}
The anomalous dimensions are
\begin{equation}\label{GNY-eta2}
 \begin{split}
  \eta_\varphi=\eta_\psi = \eta=\frac{\epsilon}{7}\,.
 \end{split}
\end{equation}

To better understand this spectrum let us consider the associated operators:
For $n=0,1,2$ the deformations $\varphi^0$, $\varphi^1$ and $\varphi^2$ do not mix
and have exponents $\theta_{n,1}$ with $n=0,1,2$.
The first mixing occurs for $n\geq 3$ among the operators $\varphi^n$ and $\varphi^{n-3}\bar{\psi}\psi$ 
and results in the $\theta_{n,i}$ exponents for $i=1,2$. The tower of operators corresponding 
to $\theta_{n,1}$ is aligned with $\varphi^n$, while
for $n\geq 3$ the tower of operators corresponding 
to $\theta_{n,2}$ is a genuine mixture of $\varphi^n$ and $\varphi^{n-3}\bar{\psi}\psi$
because the mixing matrix is triangular.

Before comparing the spectrum with \eqref{susy-spectrum} we must factor out the scaling relations and parity breaking operators.
We can prove the two scaling relations $\theta_{1,1}=(d+2-\eta)/2=3-4/7\epsilon$ and $\theta_{3,2}=(d-2+\eta)/2=1-3/7\epsilon$.
The exponent $\theta_{3,1}=1-\epsilon$ corresponds to the operator
$\varphi^3$ and, using the equations of motion, can be shown to
correspond to the parity \emph{and} supersymmetry breaking
deformation.  In Sect.~\ref{section-intro-actions}, we have associated
this to a displacement of the masses of $\varphi$ and the Majorana
excitations. Ensuring that parity is preserved, we can disregard
this deformation. Having factored out the scaling relations,
we are therefore left with two critical exponents in the relevant
part of the spectrum, $\eta_\varphi=\eta_\psi=\eta=\epsilon/7$ and
$\theta_{2,1}=\nu^{-1}=2-4/7\epsilon$. Here, we have introduced the
correlation length exponent $\nu$ corresponding to the leading
relevant deformation of the order parameter potential. We observe that $\eta$ and $\nu$ coincide
with $\eta$ and $\nu_W$ of the supersymmetric model in the previous section. This observation is key to
understanding the emergence of supersymmetry,
as we shall further explore in the next Section.

Let us also comment on the near marginal critical exponents. The mixing of the 
operators $\varphi^4$ and $\varphi\,\bar{\psi}\psi$ is expected to result in two directions: 
one that respects the constraints of supersymmetry, and another that breaks it. It is clear that
the exponents $\theta_{4,1}=-\epsilon$ corresponds to the former class as we have seen it in the supersymmetric 
spectrum, while $\theta_{4,2}=-13/7\epsilon$ corresponds to the latter. At this stage, it is an interesting observation that 
we can conjecture not only that supersymmetry is emergent in the infrared,
but also that the first irrelevant deviation from supersymmetry is subleading compared to the first irrelevant 
supersymmetric deformation! The supersymmetry breaking exponent $\tilde{\omega}\equiv -\theta_{4,2}$ 
is expected to be larger than the supersymmetric one $\omega =-\theta_{4,1}$. We 
anticipate now that it will be an important feature of our approach to reproduce this fact correctly 
also in the three dimensional case. 

As for Section \ref{susy-model-section}, let us conclude by providing the explicit form for some of the the deformations of the effective action.
As previously stated, the critical exponents $\theta_{n,1}$ are associated to deformations of the form $\varphi^n$,
therefore the scaling operators for $\theta_{3,1}$ and $\theta_{4,1}$ are purely in the bosonic potential
\begin{equation}\label{GNY-explicit-deformations1}
 \begin{split}
 \delta S \propto
  \begin{cases}
	\int_x \varphi^3 &{\rm for \quad}\theta_{3,1} = 1-\epsilon \,,\\
	\int_x \varphi^4 &{\rm for \quad}\theta_{4,1} = -\frac{13}{7}\epsilon \,.
  \end{cases}
 \end{split}
\end{equation}
The operators corresponding to the critical exponents $\theta_{3,2}$ and $\theta_{4,2}$ are instead
\begin{equation}\label{GNY-explicit-deformations2}
 \begin{split}
 \delta S \propto
 \begin{cases}
      \int_x \Bigl(\bar{\psi}\psi + g_1^\star \varphi^3 \Bigr) 			&{\rm for \quad}\theta_{3,2} = 1-\frac{3}{7}\epsilon\,,\\
      \int_x \Bigl(\varphi\bar{\psi}\psi + \frac{g_1^\star}{2} \varphi^4\Bigr) 	&{\rm for \quad}\theta_{4,2} = -\epsilon \,.
  \end{cases}
 \end{split}
\end{equation}
Using the explicit for of the fixed points \eqref{susy-fp} and \eqref{GNY-fp}, it is easy to see explicitly that the system 
\eqref{GNY-explicit-deformations2} coincides with the supersymmetric deformations \eqref{susy-explicit-deformations}.
This fact is further investigated in the next Section.
The system \eqref{GNY-explicit-deformations1} instead provides the mechanism for supersymmetry breaking of the Yukawa system
and will be reflected in the analysis of Sect.~\ref{section-susy-breaking-flow}.

\section{Relations between Yukawa and supersymmetric flows} \label{section-relations}

Now we want to explore the relation between the two RG systems \eqref{susy-flow} and 
\eqref{GNY-flow} while being guided by our approach in Sect.~\ref{section-intro-actions}.
Let us naively use \eqref{potentials_redefinition} and neglect the function $Y(\varphi)$ and the zero point energy in order to 
identify $u=(w')^2/2$ and $h=w''$. This implies the following relation among the couplings
\begin{equation} \label{couplings-relations}
 \begin{split}
 g_1 = \lambda\,,\qquad g_2= 3\lambda^2\,,
 \end{split}
\end{equation}
which is consistent with their values at criticality \eqref{susy-fp} and \eqref{GNY-fp}, and agrees with \cite{Sonoda:2011qd,Fei:2016sgs}.
It is a trivial exercise to use the explicit form of the effective potentials and \eqref{couplings-relations} 
in \eqref{susy-eta} and \eqref{GNY-eta} to prove
\begin{equation} 
 \begin{split}
 \eta = \eta_\varphi = \eta_\psi\,.
 \end{split}
\end{equation}

We can proceed further by substituting the identifications $u=(w')^2/2$ and $h=w''$ directly into the system \eqref{GNY-flow}.
The flow of the scalar potential becomes
\begin{eqnarray}\label{v-to-w-susy}
 \beta_u &=&
 w'\Bigl\{
 -2 w' +\frac{1}{2}(2+\eta)\varphi w'' +\frac{1}{2}\epsilon\left(w'-\varphi w''\right)
 +\frac{1}{16\pi^2}w'''(w'')^2+\frac{1}{32\pi^2}w'(w''')^2
 \Bigr\}\nonumber
 \\
 &\simeq&
 w'\Bigl\{
 \left(-2+\frac{\eta}{2}\right) w' +\frac{1}{2}(2+\eta)\varphi w'' +\frac{1}{2}\epsilon\left(w'-\varphi w''\right)
 +\frac{1}{16\pi^2}w'''(w'')^2
 \Bigr\} 
 \\
 &=& w'\left(k\partial_k w'\right) = k\partial_k \frac{(w')^2}{2} =\beta_{ (w')^2/2 }\,, \nonumber
\end{eqnarray}
In the second line, we have used the fact that the superpotential is cubic \emph{near} criticality
$$\frac{w'''(\varphi)^2}{16\pi^2} \simeq \frac{w'''(0)^2}{16\pi^2} =\eta $$
and therefore its third derivative can be used to complete the scaling terms of $k\partial_k w'$.
Equation \eqref{v-to-w-susy} demonstrates that the flows of the on-shell potential $u$
and that of the superpotential $w$ carry identical information near criticality.
Therefore, also the exponents governing the approach to criticality have to be the same.
We are thus lead to conclude that the correlation length exponent $\nu$
and the leading exponent of the deformation of the superpotential $\nu_W$ have to be identified,
$\nu\equiv \nu_W$, as is confirmed by the explicit calculation above.

Similarly the flow of the Yukawa function becomes in the same limit
\begin{equation}
 \begin{split}
 \beta_h &= (-1+\eta)w'' -\frac{1}{2}\epsilon\varphi w''' + \frac{1}{2}\varphi (2+\eta) w''' +\frac{1}{8\pi^2}w'' (w''')^2
 \\
 &= k\partial_k w'' - \frac{1}{16\pi^2}(w'')^2 w^{(4)}\,,
 \\
 &\simeq k\partial_k w'' = \beta_{w''}\,,
 \end{split}
\end{equation}
for which we have again used the fact that the critical superpotential is cubic
and therefore its fourth derivative is zero near criticality.
We observe that this near-criticality approximation does not affect the \emph{relevant}
operators of the physical spectrum and thus leaves the universality class intact;
however, we observe that this approximation may modify the spectra of the \emph{irrelevant} operators of the two systems.
More evidence of this fact is give in the next Section.

The above considerations prove that the critical points of the Yukawa and supersymmetric models are the same near the upper critical dimension.
The RG flow of the two models are related at least in the vicinity of the critical points,
because we explicitly used near-critical properties.
More specifically they explain
the recurrent critical exponents $\eta$ and $\nu$ in the spectra of the two models.
In fact, the relations $u=(w')^2/2$ and $h=w''$ could be interpreted as a non-linear redefinition of the couplings 
of the systems, and therefore the two systems are expected to be physically isospectral on general grounds.
As we have already seen, that the spectra must be compared once the scaling relations have been factored out,
because they depend on the operator parametrization; the use of the scaling relations is reminiscent to
the process of going on-shell which is expected to remove any parametrization dependence of the coupling space.

\section{Supersymmetry-breaking flow} \label{section-susy-breaking-flow}

In this Section we turn our attention to the renormalization of \eqref{truncation_yukawa_susy_off}.
The system presented here should be understood as the breaking of \eqref{susy-flow} of Sect.~\ref{susy-model-section}
which is induced by the term \eqref{truncation_breaking_term}.
As discussed in Sect.~\ref{section-general-rg-properties}, when the breaking function $Y(\varphi)$ is present
a generic RG step conflicts with the auxiliary nature of the field $F$ in that it creates an additional functional 
structure as shown in \eqref{rg-step-broken}. We shall handle this conflict with the scale-dependent redefinition 
of $F$ described in \eqref{F-field-flow}, which restores the auxiliary nature of $F$ at each scale, but 
changes the RG step as shown in \eqref{rg-step-broken3}.

The dimensionless superpotential is defined as in \eqref{dimensionless-superpotential},
but it is important to recall that this system potentially has different anomalous dimensions for $\varphi$ 
and the Majorana fermion. We also need to define the dimensionless counterpart of the breaking function:
\begin{equation}
\begin{split}
 w(\varphi) &= k^{-d+1} W(\varphi \, k^{d/2-1}Z^{-1/2}_\varphi)\,, \\ 
 y(\varphi) &= k^{-1} Z_\psi^{-1} Y(\varphi \, k^{d/2-1}Z^{-1/2}_\varphi)\,. 
\end{split}
\end{equation}

Despite having formulated the system using the function $y(\varphi)$,
it is convenient to give the results in terms of the function $h(\varphi)=w''(\varphi)+y(\varphi)$ rather 
than $y(\varphi)$. The RG flow is
\begin{equation}\label{broken-susy-flow}
 \begin{split}
  \beta_w  =& -3 w(\varphi) +\varphi w'(\varphi) + \epsilon \Bigl( w(\varphi)-\frac{1}{2}\varphi w'(\varphi)\Bigr)
  + \frac{\eta_\varphi}{2}\varphi w'(\varphi) +\frac{1}{3(4\pi)^2}w''(\varphi)^3 \\
  &+\int_0^\varphi {\rm d}x \,\frac{w''(x)^4-h(x)^4}{2(4\pi)^2 w'(x)}\,,\\
  \beta_h =& -h(\varphi) + \varphi h'(\varphi)- \frac{\epsilon}{2}\varphi h'(\varphi) -\eta_\psi h(\varphi) + \frac{\eta_\varphi}{2}\varphi h'(\varphi) + \frac{2}{(4\pi)^2} h(\varphi) h'(\varphi)^2\,.
 \end{split}
\end{equation}
The flow of $w(\varphi)$ is divided in two notable parts: The first line displays the supersymmetric invariant 
flow that characterizes the ${\cal N}=1$ model of \eqref{susy-flow}. The second line includes the effect of 
the scale-dependent $F$ field redefinition; in practice it contains the $\varphi$ integration of the function 
multiplying $F$ in RG step \eqref{rg-step-broken3}. For a simple illustration of why the numerator of 
the $F$ field redefinition is the difference of two terms we refer the reader to Fig.~\ref{diagrams}.
The flow of $h(\varphi)$ is equal to the flow of the same Yukawa function of the Yukawa model in \eqref{GNY-flow}, 
which is the reason why we use here the function $h(\varphi)$ instead of $y(\varphi)$.

The anomalous dimensions are
\begin{equation}\label{broken-susy-eta}
 \begin{split}
  \eta_\varphi=\frac{3}{16\pi^2}w'''(0)^2-\frac{1}{8\pi^2}h'(0)^2\,,\qquad \eta_\psi=\frac{1}{16\pi^2}h'(0)^2\,.
 \end{split}
\end{equation}
It is easy to see that if $y(\varphi)=0$ then
$h'(0)=w'''(0)$, as expected at the critical point. Then the two contributions of $\eta_\psi$ partly cancel, such that
$\eta_\varphi=\eta_\psi=\eta$ with $\eta$ being the anomalous dimension of the ${\cal N}=1$ system.

Despite its nonlocal nature, the last term of $\beta_w$ in \eqref{broken-susy-flow} is essential in establishing 
the relation of this system with the Yukawa system \eqref{GNY-flow}. As we have already seen, the beta function $\beta_h$ 
is identical in \eqref{broken-susy-flow} and \eqref{GNY-flow}, so it only remains to establish 
that we can reconstruct $\beta_u$ of \eqref{GNY-flow} from \eqref{broken-susy-flow} using the identification 
$u=\frac{(w')^2}{2}$. In order to do so, we begin by taking a derivative of $\beta_w$ with respect to the field.
This manipulation is very similar to the one shown in Sect.~\ref{section-relations}, so it will not be reproduced 
in it entirety here.
Following a logic similar to \eqref{v-to-w-susy} and including the new term, we see that
\begin{equation}\label{broken-susy-id}
 \begin{split}
  \beta_{w'}  \simeq & 
 -2 w' +\frac{1}{2}(2+\eta)\varphi w'' +\frac{1}{2}\epsilon\left(w'-\varphi w''\right)+\frac{1}{16\pi^2}w'''(w'')^2
 +\frac{1}{32\pi^2}w'(w''')^2 + \frac{(w'')^4-h^4}{2(4\pi)^2 w'}\,.
 \end{split}
\end{equation}
We can now multiply by $w'(\varphi)$ both sides and rediscover $\beta_v$ in the limit $u=\frac{(w')^2}{2}$ 
to prove that our identification is consistent
\begin{equation}\label{broken-susy-id2}
 \begin{split}
  w'\beta_{w'}  &\simeq   \beta_u\,.
 \end{split}
\end{equation}
As with the results of Sect.~\ref{section-relations}, the approximate equality means equality for the 
relevant part of the spectrum, but the two systems may differ with respect to their irrelevant operators.

With hindsight and in view of \eqref{broken-susy-id} and \eqref{broken-susy-id2}, the origin of the nonlocal contribution 
to $\beta_w$ appearing in \eqref{broken-susy-flow} is clear: it is necessary to balance out the actions 
of taking a field derivative and multiplying by $w'(\varphi)$. We find the steps leading to the identification 
between the systems \eqref{broken-susy-flow} and \eqref{GNY-flow} rather interesting, and
they serve as strong justification for the procedure described in Sect.~\ref{section-general-rg-properties}
and particularly for the scale-dependent redefinition of the field $F$.

We continue this Section providing the full analysis of the spectrum.
The critical potentials can be parametrized as
\begin{equation}
 \begin{split}
  w(\varphi)=\frac{\lambda}{3!}\varphi^3\,,\qquad h(\varphi)= (\lambda+y_1)\varphi\,,
 \end{split}
\end{equation}
where we have introduced the new Yukawa coupling $y_1$, which is related to $g_1$ as of $g_1=\lambda+y_1$.
The beta functions are
\begin{equation}
 \begin{split}
  \beta_\lambda &= -\frac{\epsilon}{2}\lambda+\frac{\lambda}{32\pi^2}(7\lambda^2-12\lambda y_1-6 y_1^2)\,,\\
  \beta_{y_1} &= -\frac{\epsilon}{2}y_1+\frac{y_1}{32\pi^2}(27\lambda^2+18\lambda y_1+4 y_1^2)\,.
 \end{split}
\end{equation}
The fixed point that we are interested in is
\begin{equation}
 \begin{split}
  \frac{\lambda^\star{}^2}{(4\pi)^2} = \frac{\epsilon}{7}\,,\qquad y^\star=0\,.
 \end{split}
\end{equation}
It is trivially seen to lie in the supersymmetric hypersurface and generalizes the one of Sect.~\ref{susy-model-section}
and matches the one of Sect.~\ref{GNY-section}.
At the fixed point, the spectrum of deformations displays some mixing
\begin{equation}
 \begin{split}
  \theta_{n,1} &= 3-n - \frac{1}{7}(3+n^2)\epsilon - 6\,\eta\, \delta_{n,3}\,, \qquad {\rm for }\quad n\geq0 \\
  &=\Bigl\{3-\frac{3}{7}\epsilon,\, 2-\frac{4}{7}\epsilon,\, 1-\epsilon,\, -\frac{18}{7}\epsilon,\,\dots\Bigr\}\,,\\
  \theta_{n,2} &= 3-n -\frac{1}{7}(n+1)\epsilon - 3\,\eta\, \delta_{n,3}\,, \qquad {\rm for }\quad n\geq2 \\
  &=\Bigl\{1-\frac{3}{7}\epsilon,\, -\epsilon, \,\dots\Bigr\}\,.
 \end{split}
\end{equation}
This spectrum should be compared with \eqref{GNY-spectrum} once all
scaling relations are factored out.  The relevant parts of the two
spectra coincide as they are characterized by the exponents
$\theta_{1,1}=\nu^{-1}=2-4/7\epsilon$ and $\eta=\epsilon/7$. The first
irrelevant exponent $\theta_{3,2}=-\omega=-\epsilon$ also coincides,
but the second irrelevant exponent $\tilde{\omega}$ differs.
This makes our observation manifest that the near-criticality approximation, denoted above
by the ``$\simeq$'' symbol, preserves the relevant part of the spectrum but may
affect the irrelevant part.

The difference can be seen also at the level of the scaling operators.
In fact, the deformations of the effective on-shell action corresponding to the critical exponents 
$\theta_{2,1}$ and $\theta_{3,1}$ are
\begin{equation}
 \begin{split}
 \delta S \propto
 \begin{cases}
	\int_x \Bigl( \lambda^\star\varphi^3\Bigr) 					&{\rm for \quad}\theta_{2,1} = 1-\epsilon \,, \\
	\int_x \Bigl(3\varphi\bar{\psi}\psi + \frac{7}{3}\lambda^\star \varphi^4\Bigr) 	&{\rm for \quad}\theta_{3,1} = -\frac{18}{7}\epsilon\,,
  \end{cases}
 \end{split}
\end{equation}
in which it is easy to see that the first operator coincides with \eqref{GNY-explicit-deformations1}, while the second one does not.
The deformations of the effective on-shell action corresponding to the critical exponents $\theta_{2,2}$ 
and $\theta_{3,2}$ are instead
\begin{equation}
 \begin{split}
 \delta S \propto
 \begin{cases}
	\int_x \Bigl(\bar{\psi}\psi + \lambda^\star \varphi^3\Bigr) 			&{\rm for \quad} \theta_{2,2} = 1-\frac{3}{7}\epsilon\,, \\
	\int_x \Bigl(\varphi\bar{\psi}\psi + \frac{\lambda^\star}{2} \varphi^4\Bigr)	& {\rm for \quad}\theta_{3,2} = -\epsilon \,.
  \end{cases}
 \end{split}
\end{equation}
which coincide with both the deformations \eqref{susy-explicit-deformations} and \eqref{GNY-explicit-deformations2},
and therefore are manifestly supersymmetric scaling operators.

\section{Conclusions} \label{section-conclusions}

We have investigated the emergence of ${\cal N}=1$ supersymmetry in
the long-range behavior of three-dimensional parity-symmetric Yukawa
systems. Our functional approach to the renormalization group confirms
that a non-trivial interacting fixed-point is within the
supersymmetric hypersurface of the general model class. The essential
new aspect of our work is that we have also access to the flow outside
the supersymmetric hypersurface. Thus, it becomes possible to answer the
question as to whether this hypersurface of higher symmetry is
attractive. We answer this question in the affirmative and find that
the supersymmetric fixed point has only one relevant direction which
is fully inside the supersymmetric hypersurface. All other perturbations
within this model class are RG irrelevant and thus suppressed in
the long-range limit. Provided that the microscopic interactions are
in the vicinity of this fixed point, the long-range observables
exhibit supersymmetry -- even if the microscopic interactions are not
supersymmetric. This is the essence of emerging supersymmetry driven
by fluctuations.

These results have been facilitated by a functional renormalization
approach that manifestly preserves supersymmetry whenever such
symmetry is realized. For this, a formulation using an auxiliary field
$F$ similar to off-shell supersymmetry is useful even for the
non-supersymmetric models. We have amended the existing supersymmetric
functional RG approach with scale-dependent field transformations of
the field $F$ in order to keep track of the symmetry status on all
scales.

In the main text, we have used the techniques of the functional
perturbative RG which are most convenient for an analytical and
transparent analysis of the RG equations, making contact with standard
$\epsilon$-expansion techniques near the upper critical dimension. On
the basis of the full functional RG (see Appendices), we can work
directly in three dimensions and obtain quantitative results of higher
numerical significance. For this, we have considered the functional flows to
leading-order in the derivative expansion (so-called LPA'),
constructed the fixed-point functions and studied perturbations in the
critical regime. In addition to a regulator of type $n=1$ in the
notation of \cite{Hellwig:2015woa}, we have also studied the $n=2$
regulator. We consider the $n=2$ results as superior, as this regulator provides for
a more pronounced gap in the propagators regularizing the IR behavior
in agreement with optimization concepts \cite{Litim:2000ci}.

Our quantitative results for the anomalous dimension $\eta$, the
correlation length exponent $\nu$ and two corrections-to-scaling
exponents $\omega$ and $\tilde\omega$ are summarized in
Tab.~\ref{tab1}. In agreement with our observations using the
functional perturbative RG, we find that the exponent $\tilde\omega$
characterizing a supersymmetry-breaking perturbation is larger than
the supersymmetry preserving one $\omega$. This implies that
supersymmetry emerges even faster than the general feature of
universality at long ranges.

\begin{table}
  \begin{tabularx}{0.8\textwidth}{| c || X | X | X | X | X | X | X |}
    \hline
			& $\cei^{\rm FRG}_{n=1}$ 	& $\cei^{\rm FRG}_{n=2}$ 	& $\epsilon^2$ 			& $\epsilon^3$ [2/1] 	& $\cei^{\rm FRG}_{\rm Yukawa}$ \cite{Vacca:2015nta} 	& CB \cite{Iliesiu:2015qra}  	\\
    \hline
    \hline
    $\eta$ 		& 0.174 			& 0.167 			& 0.184  			& 0.162  		& 0.185* 						& 0.164 			\\
    \hline
    $\nu^{-1}$ 		& 1.385 			& 1.395 			& 1.408  			& 1.419  		& 1.29  						& 1.418* 			\\
    \hline
    $\omega$ 		& 0.765 			& 0.782 			& 0.700* 			& 0.885* 		& 0.796 						& -- 				\\
    \hline
    $\tilde{\omega}$ 	& 0.809 			& 0.831 			& 0.909* 			& 1.407* 		& 1.09  						& -- 				\\
    \hline
    \end{tabularx}
    \caption{Anomalous dimension $\eta$, inverse correlation length exponent $\nu^{-1}$, and the corrections-to-scaling exponents $\omega$ (supersymmetric) and $\tilde\omega$ (non-supersymmetric). The first two columns represent our functional RG results for two different regulators $n=1,2$. For comparison, we list results from the $\epsilon$ expansion \cite{Mihaila:2017ble}, a functional RG study of the Yukawa system \cite{Vacca:2015nta} and a conformal bootstrap estimate \cite{Iliesiu:2015qra}. Numbers quoted with an asterisk are not directly given in the literature, but have been deduced by us from the available literature information, see main text for details.}
\label{tab1}
\end{table}

We also list several results taken from the literature including the
$\epsilon$ expansion up to third order \cite{Mihaila:2017ble}, a
functional RG study of the Yukawa model \cite{Vacca:2015nta}
corresponding to an on-shell formulation in the context of
supersymmetry, and estimates from the conformal bootstrap
\cite{Bashkirov:2013vya,Iliesiu:2015qra,Iliesiu:2017nrv}. The numbers quoted with an asterisk,
have not been given in these papers, but are deduced by us: for instance,
we have estimated the numbers for $\omega$ and $\tilde{\omega}$ from
the $\epsilon$ expansion \cite{Mihaila:2017ble} to second and third
order. We observe that a meaningful result for the
corrections-to-scaling exponents requires a Pad\'{e} resummation
already at order $\epsilon^2$ where we have used a [1/1]
approximant. At order $\epsilon^3$, the results for $\omega$ and
$\tilde\omega$ depend strongly on the choice of the Pad\'{e}
approximant, and only the [2/1] approximant appears to be
meaningful. This fits to the observation made in
\cite{Mihaila:2017ble} that the superscaling relation is preserved at
order $\epsilon^3$ using the [2/1] Pad{\'e} approximant.
Within the functional RG method of this paper, it can be satisfied with arbitrary accuracy.
The functional RG study of the Yukawa model in \cite{Vacca:2015nta} uses a non-supersymmetric
regularization scheme which for the $\mathcal{N}=1$ model contaminates
the anomalous dimension such that $\eta_\psi\neq\eta_\varphi$. For our
comparison, we use the arithmetic mean of their rather different
values $\eta_\varphi=0.14$ and $\eta_\psi=0.22$. From the
conformal bootstrap, only an estimate of $\eta$ is available
\cite{Bashkirov:2013vya,Iliesiu:2015qra}. We have added the estimate for the
correlation length exponent using the superscaling relation \eqref{superscaling-relation}.

We observe a comparatively mild dependence of our results on the
regularization scheme for $\eta$ and $\nu$ and a slightly bigger dependence for
$\omega$ and $\tilde\omega$, providing a naive error estimate. The
agreement with the $\epsilon$ expansion and with the conformal
bootstrap is rather satisfactory for $\eta$ and $\nu$, also being in
line with our expectation that our results for the $n=2$ regulator are
superior. The agreement is less quantitative for the subleading
exponent $\tilde\omega$. On the functional RG side, we expect more
substantial corrections from higher-order operators, while the absence
of apparent convergence and the stronger dependence of the $\epsilon$
expansion on the choice of the Pad{\'e} approximants indicates that
higher-orders are needed here as well.

In summary, we consider the understanding of the emergence of
supersymmetry in the present model class as rather comprehensive on
the qualitative side with some more room for quantitative improvement
left on the side of the corrections-to-scaling exponents.

\appendix

\section{${\cal N}=1$ and broken-supersymmetric non-perturbative flows} \label{susy-breaking-appendix}

In this Appendix we shall list the non-perturbative beta functions and anomalous dimensions for the systems 
\eqref{truncation_susy_off} and \eqref{truncation_yukawa_susy_off}, which include both the physical degrees
of freedom $\varphi$ and $\psi$ of the original system \eqref{truncation_yukawa}, as well as the auxiliary $F$ field.
These non-perturbative RG functions have been computed using the functional RG methods
formulated in terms of a flow equation for the 1PI effective action \cite{Wetterich:1992yh}.
Using a manifestly off-shell supersymmetric regularization \cite{Synatschke:2008pv,Synatschke:2009nm},
the functional RG equations for the present model have been derived and analyzed
in \cite{Mastaler:2012xg,Heilmann:2014iga} and were extended to arbitrary dimension $2<d\leq 4$ in \cite{Hellwig:2015woa}.
These methods allow for the construction of an RG which is manifestly off-shell supersymmetric at any 
scale and in any dimension as long as the supersymmetry imposed on the level of the action.
We here present all results using an optimized cutoff function known as ``$n=1$'' cutoff in the notation of \cite{Hellwig:2015woa},
and refer to this latter paper both for further details and for explicit formulas involving the ``$n=2$'' cutoff.
The results for both cutoffs are presented in Table \ref{tab1}.
We refer the reader to the discussion of Sect.~\ref{section-general-rg-properties},
which we shall follow throughout this Appendix.

The nonperturbative flow of the ${\cal N}=1$ model is
\begin{equation}\label{susy-flow-nonperturbative}
 \begin{split}
  \beta_w &= -(d-1)w+\frac{d-2+\eta}{2}\varphi w'
  -\frac{1}{2(4\pi)^{d/2}\Gamma(1+d/2)(d-1)} \frac{d-\eta}{{1+(w'')^2}} w'',
  \\
  \eta &= \frac{d (1-\frac{\eta}{d-1})}{(4\pi)^{d/2}\Gamma(1+d/2)(d-2)} \frac{1-(w'')^2}{(1+(w'')^2)^3}w'''^2.
 \end{split}
\end{equation}
This RG step corresponds to \eqref{rg-step-susy}, which is the one that preserves off-shell supersymmetry.
All the fields $\varphi$, $\psi$ and $F$ of the supermultiplet share the same anomalous dimension $\eta$.
Following the procedure described in appendix \ref{appendix-VS} of this paper and in Appendix C of \cite{Codello:2017hhh}, it is straightforward to
isolate the contributions of the RG flow \eqref{susy-flow-nonperturbative} coming from logarithmic terms in $d=4-\epsilon$,
and deduce the one loop perturbative flow \eqref{susy-flow}.

Let us reserve $\eta$ to denote the anomalous dimension of the field $F$, which can be computed from the running of its two point function,
and use $\eta_\varphi$ and $\eta_\psi$ to denote the anomalous dimensions of the two physical fields as in the main text.
Under the influence of the breaking term \eqref{truncation_breaking_term}, the RG step is modified to \eqref{rg-step-broken} 
and to \eqref{rg-step-broken3}. From the coefficient of the $F$ and $F^2$ terms we determine the corrections to 
the system \eqref{susy-flow-nonperturbative}
\begin{equation}\label{susy-flow-broken-intermediate-nonperturbative}
 \begin{split}
 \beta_w =& -(d-1)w+\frac{d-2+\eta_\varphi}{2}\varphi w'
  -\frac{1}{4(4\pi)^{d/2}\Gamma(1+d/2)(d-1)} \Bigl\{ \frac{2d(1-\frac{\eta+\eta_\varphi}{2d})}{{1+(w'')^2}}
  w''
 \\&
 \hspace{6cm} +(\eta_\varphi-\eta)\arctan{(w'')} 
 \Bigr\}+\delta\beta_w,
 \\
  \eta =&
  \frac{d}{(4\pi)^{d/2}\Gamma(1+d/2)(d-2)}\Bigl\{
   \frac{1-\frac{\eta_\varphi}{d-1}}{(1+(w'')^2)^3}w'''^2
   -\frac{1-\frac{\eta}{d-1}}{(1+(w'')^2)^3}(w'')^2 w'''^2 
  \Bigr\}\,.
 \end{split}
\end{equation}
In \eqref{susy-flow-broken-intermediate-nonperturbative}, the beta function of $w$
splits into two parts: all the terms that are shown explicitly are evaluated at fixed $F$ field, and therefore 
do not include any field redefinition as in \eqref{rg-step-broken}.
Conversely, the contribution $\delta\beta_w$ arises 
from the field redefinition of \eqref{rg-step-broken3} and is discussed below.
The breaking term generates a flow for the function $H(\varphi)$
\begin{equation}\label{susy-flow-broken-nonperturbative2}
 \begin{split}
  \beta_h =&
  -(1-\eta_\psi) h +\frac{d-2+\eta_\varphi}{2}\varphi h'
  +\frac{1}{(4\pi)^{d/2}\Gamma(d/2)(d-1)}
  \Bigl\{ \frac{4\left(1-\frac{\eta_\psi}{d}\right)h(h')^2}{(1+h^2)(1+(w'')^2)^2}
  \\
  &
  \hspace{4cm}+\left[\left(1-\frac{\eta}{d}\right)w''^2
  -\left(1-\frac{\eta_\varphi}{d}\right)\right]\frac{(1+h^2)h''-2h(h')^2}{(1+h^2)(1+(w'')^2)^2}
 \Bigr\}\,.
 \end{split}
\end{equation}
This is read off from the $\bar{\psi}\psi$ term of \eqref{rg-step-broken}, and  is consequently also given 
at fixed $F$ field. The anomalous dimension $\eta_\varphi$ and $\eta_\psi$ can be written down implicitly as
\begin{equation}\label{susy-flow-broken-nonperturbative3}
 \begin{split}
  \eta_\psi =
  &\frac{d}{(4\pi)^{d/2}\Gamma(1+d/2)(d-2)} \Bigl\{
  \frac{1-\frac{2\eta_\psi-\eta_\varphi}{d-1}+\left(1-\frac{\eta}{d-1}\right)w''}{(1+w''^2)(1+h^2)^2}
    \Bigr\}(1-h^2)h'^2
  \\
   \eta_\varphi=&
   \frac{d}{(4\pi)^{d/2}\Gamma(1+d/2)(d-2)}
   \Bigl\{
   2\left(1-\frac{\eta_\psi}{d-1}\right)\Bigl[
    \frac{1-10h^2+5h^4}{(1+h^2)^5}(h')^2
  \Bigr]
  \\
  &
  +2\frac{(3-(w'')^2)\left(\left(1-\frac{\eta}{d-1}\right)(1+(w'')^4)
  +6\left(1-\frac{\eta+2\eta_\varphi}{3(d-1)}\right)(w'')^2\right)}{(1+(w'')^2)^5}(w''')^2
  \Bigr\}\,.
 \end{split}
\end{equation}
Isolating the logarithmic contributions from formulas \eqref{susy-flow-broken-intermediate-nonperturbative}, \eqref{susy-flow-broken-nonperturbative2} and  \eqref{susy-flow-broken-nonperturbative3}
one can isolate the leading one loop perturbative flow \eqref{broken-susy-flow} and \eqref{broken-susy-eta}.

As discussed in Sect.~\ref{section-general-rg-properties}, the breaking term \eqref{truncation_breaking_term} 
generates the contribution $C(\varphi)$ in the RG step \eqref{rg-step-broken} which is not present in the 
manifestly supersymmetric step \eqref{rg-step-susy}. The dimensionless counterpart $c(\varphi)$ of this function 
can be considered as a beta function and perturbatively describes the generation of an additional $\varphi^4$-like 
interaction through the diagrams of Fig.~\ref{diagrams}. We compute it as
\begin{equation}
 \begin{split}
  c(\varphi)=&\frac{1}{(4\pi)^{d/2}\Gamma(1+d/2)}\Bigl\{\frac{1-\frac{\eta+\eta_\varphi}{2(d+1)}}{1+(w'')^2}
  -\frac{1-\frac{\eta_\psi}{d+1}}{1+h^2}\Bigr\}\,.
 \end{split}
\end{equation}
Following the reasoning of Sect.~\ref{section-general-rg-properties}, the generation of this self-interaction can be compensated by a scale-dependent redefinition of the
$F$ field along the flow 
for each RG step \eqref{rg-step-broken2}. The dimensionless beta function of the $F$ field redefinition is then 
\begin{equation}
 \begin{split}
  \beta_F &= c(\varphi)/w'(\varphi)\,.
 \end{split}
\end{equation}

When including the beta function of the $F$ redefinition it is important to carefully treat potential singularities 
that are due to the presence of the inverse power of $w'(\varphi)$. This typically means subtracting a 
double pole in $\varphi=0$ when considering the perturbative flow (because $w'\sim\varphi^2$), or a single 
pole when considering the non-perturbative one (because the minimum of the on-shell potential is non-zero). 
In practice, this means that in the computation of the spectrum the fixed point limit has to be taken carefully
and paying special attention to the first two couplings of the expansion of $w(\varphi)$.
Recall from the discussion of Sect.~\ref{section-intro-actions} that the zeroes of $w'(\varphi)$ are in one-to-one 
correspondence with the minima of the on-shell potential.
The redefinition of $F$ enters linearly in the flow of $w'(\varphi)$ according to \eqref{rg-step-broken2}, therefore 
it contributes to the flow of $w(\varphi)$ as an integral
\begin{equation}
 \begin{split}
  \delta \beta_w &= -\int^\varphi_{\varphi_0} {\rm d}x \frac{c(x)}{w'(x)}\,.
 \end{split}
\end{equation}
Our convention is to choose the boundary of the integration to be $\varphi_0$, so that the contribution $\delta \beta_w$
does not affect the location of the minimum (even though the remainder of $\beta_w$ does).
With this convention we ensure that the only change in the scaling properties of the zero point energy can 
come from $\beta_{U_0}$ of \eqref{rg-step-broken}, which we decided to neglect in the first place. If a zero 
point energy were to be reinstated as $\beta_{U_0}$, even though it would serve as an order parameter for 
supersymmetry breaking, it would only contribute to our spectra by including the critical exponent $4-\epsilon$ 
whenever it is missing.

\section{Non-perturbative flow of the Yukawa system}  \label{GNY-appendix}

The non-perturbative beta functions of the two potentials appearing in a truncation of the effective average 
action of the form \eqref{truncation_yukawa} have been considered in \cite{Zanusso:2009bs,Vacca:2010mj,Zanusso:PhD} 
and studied extensively in \cite{Vacca:2015nta} for arbitrary number and type of spinors and in general $d$ dimensions,
see also \cite{Hofling:2002hj,Braun:2010tt,Mesterhazy:2012ei,Jakovac:2014lqa,Borchardt:2015rxa,Knorr:2016sfs}.
We are interested in the case involving one Majorana fermion in $d=3$, which can be achieved by 
considering $N_f=1/4$ Dirac spinors in $d=4$.
In the notation of \cite{Vacca:2015nta}, this latter limit corresponds to set $X_f=1$, because $X_f$ counts the 
number of fermionic degrees of freedom. Using an optimized cutoff \cite{Litim:2000ci,Litim:2001up}, the beta functionals are
\begin{equation}\label{GNY-flow-nonperturbative}
 \begin{split}
  \beta_u =&
  -du + \frac{1}{2}(d-2+\eta_\varphi)\varphi u'
  +\frac{1}{(4\pi)^{d/2}\Gamma(1+d/2)}\Bigl\{
  \frac{1-\frac{\eta_\varphi}{d+2}}{1+u''}
  -\frac{1-\frac{\eta_\psi}{d+1}}{1+h^2}
  \Bigr\}
  \,,\\
  \beta_h =& -(1-\eta_\psi)h + \frac{1}{2}(d-2+\eta_\varphi)\varphi h'
  +\frac{1}{(4\pi)^{d/2}\Gamma(1+d/2)}\Bigl\{
   \frac{2\left(1-\frac{\eta_\varphi}{d+2}\right)(h')^2 h }{(1+u'')^2(1+h^2)}
   \\
   &
   +\frac{2\left(1-\frac{\eta_\psi}{d+1}\right)(h')^2 h }{(1+u'')(1+h^2)^2}
   -\frac{1-\frac{\eta_\varphi}{d+2}}{(1+u'')^2}h''
  \Bigr\}\,.
 \end{split}
\end{equation}
The anomalous dimensions of scalar and Majorana fields can be expressed implicitly as
\begin{equation}\label{GNY-eta-nonperturbative}
 \begin{split}
  \eta_\varphi=
  & \frac{1}{(4\pi)^{d/2}\Gamma(1+d/2)}\Bigl\{
  \frac{2}{(1+h^2)^4}-\frac{h^2}{(1+h^2)^4}+\frac{1-\eta_\psi}{(d-2)(1+h^2)^3}-\frac{\frac{1}{2}
  +\frac{1-\eta_\psi}{(d-2)}}{2(1+h^2)^2}
  \Bigr\}(h')^2
  \\
  & +\frac{1}{(4\pi)^{d/2}\Gamma(1+d/2)}\frac{(u''')^2}{(1+u'')^4}\,,
  \\
  \eta_\psi =
  & \frac{1}{(4\pi)^{d/2}\Gamma(1+d/2)} \frac{2\left(1-\frac{\eta_\psi}{d+1}\right)(h')^2}{(1+h^2)(1+u'')^2}\,.
 \end{split}
\end{equation}
The above RG system depends on the chosen cutoff
through the several threshold functions that parametrize the decoupling of higher modes from the flow and characterize the cutoff scheme dependence.
There are, however, some key contributions to the flow which do not depend by the cutoff in $d=4$, and are thus universal.
More specifically, all the universal one loop contributions contained in \eqref{GNY-flow-nonperturbative} 
and \eqref{GNY-eta-nonperturbative} have been given in \eqref{GNY-flow} and \eqref{GNY-eta} and 
have been the main tool to illustrate the results of Sect.~\ref{GNY-section}. It is a straightforward computation 
to check the universal flow of of Sect.~\ref{GNY-section} from \eqref{GNY-flow-nonperturbative} 
and \eqref{GNY-eta-nonperturbative} in $d=4-\epsilon$. A more detailed discussion on the role of universality in functional 
renormalization and its relation with perturbation theory can be found in \cite{Codello:2017hhh}.

\section{On the perturbative RG}
\label{appendix-VS}

The functional perturbative flows for the potentials that are used in the main text can all be obtained
by applying the methods of \cite{ODwyer:2007brp,Codello:2017hhh} to renormalize the bare actions shown in Sect.~\ref{section-intro-actions}.
As discussed in \cite{Codello:2017hhh}, the functional perturbative RG methods
fully reproduce the standard methods of perturbation theory,
but also allows the determination of additional important quantities, such as
the coefficients of some of the operator product expansion of the critical quantum field theory.
The non-perturbative flows given in appendices \ref{susy-breaking-appendix} and \ref{GNY-appendix}
seem very different from their perturbative counterparts that appeared in the main text,
but we can use the former to derive the latter. This is possible, since, even within our truncation,
the non-perturbative flows contain all the universal leading perturbative contributions;
for the examples of this paper, these occur at one-loop order.
%
%
In this appendix, we give a simple procedure to determine the one-loop perturbative \emph{universal}
RG flow starting from the non-perturbative ones of appendices \ref{susy-breaking-appendix} and \ref{GNY-appendix}.
We illustrate the procedure using the flow of the ${\cal N}=1$ superpotential.

Recalling the relation \eqref{dimensionless-superpotential} between the dimensionless superpotential $w(\varphi)$
and the dimensionful bare superpotential $W$ that appears in \eqref{truncation_susy_off},
it is easy to use \eqref{susy-flow-nonperturbative} to reconstruct the non-perturbative flow
of the \emph{dimensionful} superpotential
\begin{equation}
 \begin{split}
  \beta_W &=  -\frac{Z (d-\eta ) }{2(4\pi)^{d/2}\Gamma(1+d/2)(d-1) }\frac{k^d \, W''}{k^2 Z^2+(W^{\prime\prime })^2}\,, \\
  \eta &= \frac{Z d (1-\frac{\eta}{d-1})}{(4\pi)^{d/2}\Gamma(1+d/2)(d-2)} \frac{k^2 Z^2-(W'')^2}{(k^2 Z^2+(W'')^2)^3}(W''')^2\,.
 \end{split}
\end{equation}
We are interested in isolating the logarithmic scaling terms,
which are known to correspond to the $1/\epsilon$ poles of dimensional regularization
and which are responsible for the RG flow in the $\overline{\rm MS}$ scheme.
This scaling analysis can be simplified by first expanding the flow in powers of $W^{\prime\prime}$.
We begin by expanding $\beta_W$,
\begin{equation}\label{expanded-beta-W-dimful}
 \begin{split}
  \beta_W
  &=
  -\frac{d}{(4\pi)^{d/2}\Gamma(1+d/2)} \frac{ W''}{2 (d-1) Z }\,k^{d-2}
  +\frac{d}{(4\pi)^{d/2}\Gamma(1+d/2)}\frac{(W'')^3}{2 (d-1) Z^3 }\, k^{d-4}
  +{\cal O}(k^{d-6})\,.
 \end{split}
\end{equation}
The same procedure can be applied to $\eta$
\begin{equation}\label{expanded-eta-dimful}
 \begin{split}
  \eta &= \frac{d}{(4\pi)^{d/2}\Gamma(1+d/2)}\frac{(W''')^2}{(d-2) Z^3 }\, k^{d-4}
  +{\cal O}(k^{d-6})\,.
 \end{split}
\end{equation}

The logarithmic scaling in $d=4$ corresponds to the second term of \eqref{expanded-beta-W-dimful}
and the first of \eqref{expanded-eta-dimful}. If integrated over the scale $k$ in $d=4$,
these terms produce the familiar logarithmic singularities
$$
 \int^{\Lambda}_0 \frac{{\rm d}k}{k} \,k^{d-4} \sim \log \Lambda.
$$
These are known to be in one-to-one correspondence with the $1/\epsilon$ poles of dimensional regularization.
The four dimensional $\overline{\rm MS}$ perturbative flow of the superpotential and the anomalous dimension
can be obtained by specializing to $d=4$ and \emph{dropping} all scaling terms with the exception of the logarithmic ones
(this step removes both infrared and ultraviolet relevant contributions).
The result is
\begin{equation}
 \begin{split}
  \beta_W
  =
  \frac{1}{3(4\pi)^{2}}\frac{(W'')^3}{Z^3 }\,, \qquad \eta = \frac{1}{(4\pi)^{d/2}}\frac{(W''')^2}{ Z^3 }\,.
 \end{split}
\end{equation}
Returning to dimensionless renormalized quantities it is easy to use this latter result to prove
the perturbative flow given in \eqref{susy-flow} with which we illustrated the results of Sect.~\ref{susy-model-section}.

An interesting final point to discuss involves the first term of \eqref{expanded-beta-W-dimful}
which describes a logarithmic scaling in $d=2$. Using the same reasoning as in the last appendix of \cite{Codello:2017hhh},
the corresponding RG system appears to describe
an ${\cal N}=1$ supersymmetric generalization of the Sine-Gordon universality class.
We hope to return to this topic with a more detailed discussion in the future.


\acknowledgments

We authors are grateful to M.~M.~Scherer and L.~Zambelli for discussions.
OZ thanks A.~Codello, M.~Safari and G.~P.~Vacca for discussions and collaborations on related topics.
This work was supported by the DFG Research Training Unit GRK 1523/2.
HG and OZ acknowledge support by the DFG under Grant No.~Gi 328/6-2 and
Gi 328/7-1. AW acknowledges support by the DFG under Grant No.~Wi 777/11-1.

\bibliography{bibliography}

\begin{thebibliography}{68}%
\makeatletter
\providecommand \@ifxundefined [1]{%
 \@ifx{#1\undefined}
}%
\providecommand \@ifnum [1]{%
 \ifnum #1\expandafter \@firstoftwo
 \else \expandafter \@secondoftwo
 \fi
}%
\providecommand \@ifx [1]{%
 \ifx #1\expandafter \@firstoftwo
 \else \expandafter \@secondoftwo
 \fi
}%
\providecommand \natexlab [1]{#1}%
\providecommand \enquote  [1]{``#1''}%
\providecommand \bibnamefont  [1]{#1}%
\providecommand \bibfnamefont [1]{#1}%
\providecommand \citenamefont [1]{#1}%
\providecommand \href@noop [0]{\@secondoftwo}%
\providecommand \href [0]{\begingroup \@sanitize@url \@href}%
\providecommand \@href[1]{\@@startlink{#1}\@@href}%
\providecommand \@@href[1]{\endgroup#1\@@endlink}%
\providecommand \@sanitize@url [0]{\catcode `\\12\catcode `\$12\catcode
  `\&12\catcode `\#12\catcode `\^12\catcode `\_12\catcode `\%12\relax}%
\providecommand \@@startlink[1]{}%
\providecommand \@@endlink[0]{}%
\providecommand \url  [0]{\begingroup\@sanitize@url \@url }%
\providecommand \@url [1]{\endgroup\@href {#1}{\urlprefix }}%
\providecommand \urlprefix  [0]{URL }%
\providecommand \Eprint [0]{\href }%
\providecommand \doibase [0]{http://dx.doi.org/}%
\providecommand \selectlanguage [0]{\@gobble}%
\providecommand \bibinfo  [0]{\@secondoftwo}%
\providecommand \bibfield  [0]{\@secondoftwo}%
\providecommand \translation [1]{[#1]}%
\providecommand \BibitemOpen [0]{}%
\providecommand \bibitemStop [0]{}%
\providecommand \bibitemNoStop [0]{.\EOS\space}%
\providecommand \EOS [0]{\spacefactor3000\relax}%
\providecommand \BibitemShut  [1]{\csname bibitem#1\endcsname}%
\let\auto@bib@innerbib\@empty
\bibitem [{\citenamefont {Chadha}\ and\ \citenamefont
  {Nielsen}(1983)}]{Chadha:1982qq}%
  \BibitemOpen
  \bibfield  {author} {\bibinfo {author} {\bibfnamefont {S.}~\bibnamefont
  {Chadha}}\ and\ \bibinfo {author} {\bibfnamefont {H.~B.}\ \bibnamefont
  {Nielsen}},\ }\href {\doibase 10.1016/0550-3213(83)90081-0} {\bibfield
  {journal} {\bibinfo  {journal} {Nucl. Phys.}\ }\textbf {\bibinfo {volume}
  {B217}},\ \bibinfo {pages} {125} (\bibinfo {year} {1983})}\BibitemShut
  {NoStop}%
\bibitem [{\citenamefont {Herbut}(2006)}]{Herbut:2006cs}%
  \BibitemOpen
  \bibfield  {author} {\bibinfo {author} {\bibfnamefont {I.~F.}\ \bibnamefont
  {Herbut}},\ }\href {\doibase 10.1103/PhysRevLett.97.146401} {\bibfield
  {journal} {\bibinfo  {journal} {Phys. Rev. Lett.}\ }\textbf {\bibinfo
  {volume} {97}},\ \bibinfo {pages} {146401} (\bibinfo {year} {2006})},\
  \Eprint {http://arxiv.org/abs/cond-mat/0606195} {arXiv:cond-mat/0606195
  [cond-mat]} \BibitemShut {NoStop}%
\bibitem [{\citenamefont {Juricic}\ \emph {et~al.}(2009)\citenamefont
  {Juricic}, \citenamefont {Herbut},\ and\ \citenamefont
  {Semenoff}}]{Juricic:2009px}%
  \BibitemOpen
  \bibfield  {author} {\bibinfo {author} {\bibfnamefont {V.}~\bibnamefont
  {Juricic}}, \bibinfo {author} {\bibfnamefont {I.~F.}\ \bibnamefont {Herbut}},
  \ and\ \bibinfo {author} {\bibfnamefont {G.~W.}\ \bibnamefont {Semenoff}},\
  }\href {\doibase 10.1103/PhysRevB.80.081405} {\bibfield  {journal} {\bibinfo
  {journal} {Phys. Rev.}\ }\textbf {\bibinfo {volume} {B80}},\ \bibinfo {pages}
  {081405} (\bibinfo {year} {2009})},\ \Eprint {http://arxiv.org/abs/0906.3513}
  {arXiv:0906.3513 [cond-mat.str-el]} \BibitemShut {NoStop}%
\bibitem [{\citenamefont {Herbut}\ \emph {et~al.}(2009)\citenamefont {Herbut},
  \citenamefont {Juricic},\ and\ \citenamefont {Vafek}}]{Herbut:2009vu}%
  \BibitemOpen
  \bibfield  {author} {\bibinfo {author} {\bibfnamefont {I.~F.}\ \bibnamefont
  {Herbut}}, \bibinfo {author} {\bibfnamefont {V.}~\bibnamefont {Juricic}}, \
  and\ \bibinfo {author} {\bibfnamefont {O.}~\bibnamefont {Vafek}},\ }\href
  {\doibase 10.1103/PhysRevB.80.075432} {\bibfield  {journal} {\bibinfo
  {journal} {Phys. Rev.}\ }\textbf {\bibinfo {volume} {B80}},\ \bibinfo {pages}
  {075432} (\bibinfo {year} {2009})},\ \Eprint {http://arxiv.org/abs/0904.1019}
  {arXiv:0904.1019 [cond-mat.str-el]} \BibitemShut {NoStop}%
\bibitem [{\citenamefont {Roy}(2011)}]{Roy:PhysRevB.84.113404}%
  \BibitemOpen
  \bibfield  {author} {\bibinfo {author} {\bibfnamefont {B.}~\bibnamefont
  {Roy}},\ }\href {\doibase 10.1103/PhysRevB.84.113404} {\bibfield  {journal}
  {\bibinfo  {journal} {Phys. Rev. B}\ }\textbf {\bibinfo {volume} {84}},\
  \bibinfo {pages} {113404} (\bibinfo {year} {2011})}\BibitemShut {NoStop}%
\bibitem [{\citenamefont {Calabrese}\ \emph {et~al.}(2003)\citenamefont
  {Calabrese}, \citenamefont {Pelissetto},\ and\ \citenamefont
  {Vicari}}]{Calabrese:2002bm}%
  \BibitemOpen
  \bibfield  {author} {\bibinfo {author} {\bibfnamefont {P.}~\bibnamefont
  {Calabrese}}, \bibinfo {author} {\bibfnamefont {A.}~\bibnamefont
  {Pelissetto}}, \ and\ \bibinfo {author} {\bibfnamefont {E.}~\bibnamefont
  {Vicari}},\ }\href {\doibase 10.1103/PhysRevB.67.054505} {\bibfield
  {journal} {\bibinfo  {journal} {Phys. Rev.}\ }\textbf {\bibinfo {volume}
  {B67}},\ \bibinfo {pages} {054505} (\bibinfo {year} {2003})},\ \Eprint
  {http://arxiv.org/abs/cond-mat/0209580} {arXiv:cond-mat/0209580 [cond-mat]}
  \BibitemShut {NoStop}%
\bibitem [{\citenamefont {Eichhorn}\ \emph {et~al.}(2013)\citenamefont
  {Eichhorn}, \citenamefont {Mesterh\'{a}zy},\ and\ \citenamefont
  {Scherer}}]{Eichhorn:2013zza}%
  \BibitemOpen
  \bibfield  {author} {\bibinfo {author} {\bibfnamefont {A.}~\bibnamefont
  {Eichhorn}}, \bibinfo {author} {\bibfnamefont {D.}~\bibnamefont
  {Mesterh\'{a}zy}}, \ and\ \bibinfo {author} {\bibfnamefont {M.~M.}\
  \bibnamefont {Scherer}},\ }\href {\doibase 10.1103/PhysRevE.88.042141}
  {\bibfield  {journal} {\bibinfo  {journal} {Phys. Rev.}\ }\textbf {\bibinfo
  {volume} {E88}},\ \bibinfo {pages} {042141} (\bibinfo {year} {2013})},\
  \Eprint {http://arxiv.org/abs/1306.2952} {arXiv:1306.2952
  [cond-mat.stat-mech]} \BibitemShut {NoStop}%
\bibitem [{\citenamefont {Gomes}(2016)}]{Gomes:2015eea}%
  \BibitemOpen
  \bibfield  {author} {\bibinfo {author} {\bibfnamefont {P.~R.~S.}\
  \bibnamefont {Gomes}},\ }\href {\doibase 10.1142/S0217751X1630009X}
  {\bibfield  {journal} {\bibinfo  {journal} {Int. J. Mod. Phys.}\ }\textbf
  {\bibinfo {volume} {A31}},\ \bibinfo {pages} {1630009} (\bibinfo {year}
  {2016})},\ \Eprint {http://arxiv.org/abs/1510.04492} {arXiv:1510.04492
  [hep-th]} \BibitemShut {NoStop}%
\bibitem [{\citenamefont {Lee}(2007)}]{Lee:2006if}%
  \BibitemOpen
  \bibfield  {author} {\bibinfo {author} {\bibfnamefont {S.-S.}\ \bibnamefont
  {Lee}},\ }\href {\doibase 10.1103/PhysRevB.76.075103} {\bibfield  {journal}
  {\bibinfo  {journal} {Phys. Rev.}\ }\textbf {\bibinfo {volume} {B76}},\
  \bibinfo {pages} {075103} (\bibinfo {year} {2007})},\ \Eprint
  {http://arxiv.org/abs/cond-mat/0611658} {arXiv:cond-mat/0611658 [cond-mat]}
  \BibitemShut {NoStop}%
\bibitem [{\citenamefont {Grover}\ \emph {et~al.}(2014)\citenamefont {Grover},
  \citenamefont {Sheng},\ and\ \citenamefont {Vishwanath}}]{Grover:2013rc}%
  \BibitemOpen
  \bibfield  {author} {\bibinfo {author} {\bibfnamefont {T.}~\bibnamefont
  {Grover}}, \bibinfo {author} {\bibfnamefont {D.~N.}\ \bibnamefont {Sheng}}, \
  and\ \bibinfo {author} {\bibfnamefont {A.}~\bibnamefont {Vishwanath}},\
  }\href {\doibase 10.1126/science.1248253} {\bibfield  {journal} {\bibinfo
  {journal} {Science}\ }\textbf {\bibinfo {volume} {344}},\ \bibinfo {pages}
  {280} (\bibinfo {year} {2014})},\ \Eprint {http://arxiv.org/abs/1301.7449}
  {arXiv:1301.7449 [cond-mat.str-el]} \BibitemShut {NoStop}%
\bibitem [{\citenamefont {Ponte}\ and\ \citenamefont
  {Lee}(2014)}]{Ponte:2012ru}%
  \BibitemOpen
  \bibfield  {author} {\bibinfo {author} {\bibfnamefont {P.}~\bibnamefont
  {Ponte}}\ and\ \bibinfo {author} {\bibfnamefont {S.-S.}\ \bibnamefont
  {Lee}},\ }\href {\doibase 10.1088/1367-2630/16/1/013044} {\bibfield
  {journal} {\bibinfo  {journal} {New J. Phys.}\ }\textbf {\bibinfo {volume}
  {16}},\ \bibinfo {pages} {013044} (\bibinfo {year} {2014})},\ \Eprint
  {http://arxiv.org/abs/1206.2340} {arXiv:1206.2340 [cond-mat.str-el]}
  \BibitemShut {NoStop}%
\bibitem [{\citenamefont {Jian}\ \emph {et~al.}(2015)\citenamefont {Jian},
  \citenamefont {Jiang},\ and\ \citenamefont {Yao}}]{Jian:2014pca}%
  \BibitemOpen
  \bibfield  {author} {\bibinfo {author} {\bibfnamefont {S.-K.}\ \bibnamefont
  {Jian}}, \bibinfo {author} {\bibfnamefont {Y.-F.}\ \bibnamefont {Jiang}}, \
  and\ \bibinfo {author} {\bibfnamefont {H.}~\bibnamefont {Yao}},\ }\href
  {\doibase 10.1103/PhysRevLett.114.237001} {\bibfield  {journal} {\bibinfo
  {journal} {Phys. Rev. Lett.}\ }\textbf {\bibinfo {volume} {114}},\ \bibinfo
  {pages} {237001} (\bibinfo {year} {2015})},\ \Eprint
  {http://arxiv.org/abs/1407.4497} {arXiv:1407.4497 [cond-mat.str-el]}
  \BibitemShut {NoStop}%
\bibitem [{\citenamefont {Goh}\ \emph {et~al.}(2005)\citenamefont {Goh},
  \citenamefont {Luty},\ and\ \citenamefont {Ng}}]{Goh:2003yr}%
  \BibitemOpen
  \bibfield  {author} {\bibinfo {author} {\bibfnamefont {H.-S.}\ \bibnamefont
  {Goh}}, \bibinfo {author} {\bibfnamefont {M.~A.}\ \bibnamefont {Luty}}, \
  and\ \bibinfo {author} {\bibfnamefont {S.-P.}\ \bibnamefont {Ng}},\ }\href
  {\doibase 10.1088/1126-6708/2005/01/040} {\bibfield  {journal} {\bibinfo
  {journal} {JHEP}\ }\textbf {\bibinfo {volume} {01}},\ \bibinfo {pages} {040}
  (\bibinfo {year} {2005})},\ \Eprint {http://arxiv.org/abs/hep-th/0309103}
  {arXiv:hep-th/0309103 [hep-th]} \BibitemShut {NoStop}%
\bibitem [{\citenamefont {Antipin}\ \emph {et~al.}(2013)\citenamefont
  {Antipin}, \citenamefont {Mojaza}, \citenamefont {Pica},\ and\ \citenamefont
  {Sannino}}]{Antipin:2011ny}%
  \BibitemOpen
  \bibfield  {author} {\bibinfo {author} {\bibfnamefont {O.}~\bibnamefont
  {Antipin}}, \bibinfo {author} {\bibfnamefont {M.}~\bibnamefont {Mojaza}},
  \bibinfo {author} {\bibfnamefont {C.}~\bibnamefont {Pica}}, \ and\ \bibinfo
  {author} {\bibfnamefont {F.}~\bibnamefont {Sannino}},\ }\href {\doibase
  10.1007/JHEP06(2013)037} {\bibfield  {journal} {\bibinfo  {journal} {JHEP}\
  }\textbf {\bibinfo {volume} {06}},\ \bibinfo {pages} {037} (\bibinfo {year}
  {2013})},\ \Eprint {http://arxiv.org/abs/1105.1510} {arXiv:1105.1510
  [hep-th]} \BibitemShut {NoStop}%
\bibitem [{\citenamefont {Sonoda}(2011)}]{Sonoda:2011qd}%
  \BibitemOpen
  \bibfield  {author} {\bibinfo {author} {\bibfnamefont {H.}~\bibnamefont
  {Sonoda}},\ }\href {\doibase 10.1143/PTP.126.57} {\bibfield  {journal}
  {\bibinfo  {journal} {Prog. Theor. Phys.}\ }\textbf {\bibinfo {volume}
  {126}},\ \bibinfo {pages} {57} (\bibinfo {year} {2011})},\ \Eprint
  {http://arxiv.org/abs/1102.3974} {arXiv:1102.3974 [hep-th]} \BibitemShut
  {NoStop}%
\bibitem [{\citenamefont {Gracey}(1990)}]{Gracey:1990sx}%
  \BibitemOpen
  \bibfield  {author} {\bibinfo {author} {\bibfnamefont {J.~A.}\ \bibnamefont
  {Gracey}},\ }\href {\doibase 10.1016/0550-3213(90)90186-H} {\bibfield
  {journal} {\bibinfo  {journal} {Nucl. Phys.}\ }\textbf {\bibinfo {volume}
  {B341}},\ \bibinfo {pages} {403} (\bibinfo {year} {1990})}\BibitemShut
  {NoStop}%
\bibitem [{\citenamefont {Hands}\ \emph {et~al.}(1993)\citenamefont {Hands},
  \citenamefont {Kocic},\ and\ \citenamefont {Kogut}}]{Hands:1992be}%
  \BibitemOpen
  \bibfield  {author} {\bibinfo {author} {\bibfnamefont {S.}~\bibnamefont
  {Hands}}, \bibinfo {author} {\bibfnamefont {A.}~\bibnamefont {Kocic}}, \ and\
  \bibinfo {author} {\bibfnamefont {J.~B.}\ \bibnamefont {Kogut}},\ }\href
  {\doibase 10.1006/aphy.1993.1039} {\bibfield  {journal} {\bibinfo  {journal}
  {Annals Phys.}\ }\textbf {\bibinfo {volume} {224}},\ \bibinfo {pages} {29}
  (\bibinfo {year} {1993})},\ \Eprint {http://arxiv.org/abs/hep-lat/9208022}
  {arXiv:hep-lat/9208022 [hep-lat]} \BibitemShut {NoStop}%
\bibitem [{\citenamefont {Vasiliev}\ \emph {et~al.}(1993)\citenamefont
  {Vasiliev}, \citenamefont {Derkachov}, \citenamefont {Kivel},\ and\
  \citenamefont {Stepanenko}}]{Vasiliev:1992wr}%
  \BibitemOpen
  \bibfield  {author} {\bibinfo {author} {\bibfnamefont {A.~N.}\ \bibnamefont
  {Vasiliev}}, \bibinfo {author} {\bibfnamefont {S.~E.}\ \bibnamefont
  {Derkachov}}, \bibinfo {author} {\bibfnamefont {N.~A.}\ \bibnamefont
  {Kivel}}, \ and\ \bibinfo {author} {\bibfnamefont {A.~S.}\ \bibnamefont
  {Stepanenko}},\ }\href {\doibase 10.1007/BF01019324} {\bibfield  {journal}
  {\bibinfo  {journal} {Theor. Math. Phys.}\ }\textbf {\bibinfo {volume}
  {94}},\ \bibinfo {pages} {127} (\bibinfo {year} {1993})},\ \bibinfo {note}
  {[Teor. Mat. Fiz.94,179(1993)]}\BibitemShut {NoStop}%
\bibitem [{\citenamefont {Rosenstein}\ \emph {et~al.}(1993)\citenamefont
  {Rosenstein}, \citenamefont {Yu},\ and\ \citenamefont
  {Kovner}}]{Rosenstein:1993zf}%
  \BibitemOpen
  \bibfield  {author} {\bibinfo {author} {\bibfnamefont {B.}~\bibnamefont
  {Rosenstein}}, \bibinfo {author} {\bibfnamefont {H.-L.}\ \bibnamefont {Yu}},
  \ and\ \bibinfo {author} {\bibfnamefont {A.}~\bibnamefont {Kovner}},\ }\href
  {\doibase 10.1016/0370-2693(93)91253-J} {\bibfield  {journal} {\bibinfo
  {journal} {Phys. Lett.}\ }\textbf {\bibinfo {volume} {B314}},\ \bibinfo
  {pages} {381} (\bibinfo {year} {1993})}\BibitemShut {NoStop}%
\bibitem [{\citenamefont {Moshe}\ and\ \citenamefont
  {Zinn-Justin}(2003)}]{Moshe:2003xn}%
  \BibitemOpen
  \bibfield  {author} {\bibinfo {author} {\bibfnamefont {M.}~\bibnamefont
  {Moshe}}\ and\ \bibinfo {author} {\bibfnamefont {J.}~\bibnamefont
  {Zinn-Justin}},\ }\href {\doibase 10.1016/S0370-1573(03)00263-1} {\bibfield
  {journal} {\bibinfo  {journal} {Phys. Rept.}\ }\textbf {\bibinfo {volume}
  {385}},\ \bibinfo {pages} {69} (\bibinfo {year} {2003})},\ \Eprint
  {http://arxiv.org/abs/hep-th/0306133} {arXiv:hep-th/0306133 [hep-th]}
  \BibitemShut {NoStop}%
\bibitem [{\citenamefont {Fei}\ \emph {et~al.}(2016)\citenamefont {Fei},
  \citenamefont {Giombi}, \citenamefont {Klebanov},\ and\ \citenamefont
  {Tarnopolsky}}]{Fei:2016sgs}%
  \BibitemOpen
  \bibfield  {author} {\bibinfo {author} {\bibfnamefont {L.}~\bibnamefont
  {Fei}}, \bibinfo {author} {\bibfnamefont {S.}~\bibnamefont {Giombi}},
  \bibinfo {author} {\bibfnamefont {I.~R.}\ \bibnamefont {Klebanov}}, \ and\
  \bibinfo {author} {\bibfnamefont {G.}~\bibnamefont {Tarnopolsky}},\ }\href
  {\doibase 10.1093/ptep/ptw120} {\bibfield  {journal} {\bibinfo  {journal}
  {PTEP}\ }\textbf {\bibinfo {volume} {2016}},\ \bibinfo {pages} {12C105}
  (\bibinfo {year} {2016})},\ \Eprint {http://arxiv.org/abs/1607.05316}
  {arXiv:1607.05316 [hep-th]} \BibitemShut {NoStop}%
\bibitem [{\citenamefont {Gracey}\ \emph {et~al.}(2016)\citenamefont {Gracey},
  \citenamefont {Luthe},\ and\ \citenamefont {Schroder}}]{Gracey:2016mio}%
  \BibitemOpen
  \bibfield  {author} {\bibinfo {author} {\bibfnamefont {J.~A.}\ \bibnamefont
  {Gracey}}, \bibinfo {author} {\bibfnamefont {T.}~\bibnamefont {Luthe}}, \
  and\ \bibinfo {author} {\bibfnamefont {Y.}~\bibnamefont {Schroder}},\ }\href
  {\doibase 10.1103/PhysRevD.94.125028} {\bibfield  {journal} {\bibinfo
  {journal} {Phys. Rev.}\ }\textbf {\bibinfo {volume} {D94}},\ \bibinfo {pages}
  {125028} (\bibinfo {year} {2016})},\ \Eprint
  {http://arxiv.org/abs/1609.05071} {arXiv:1609.05071 [hep-th]} \BibitemShut
  {NoStop}%
\bibitem [{\citenamefont {Zerf}\ \emph {et~al.}(2016)\citenamefont {Zerf},
  \citenamefont {Lin},\ and\ \citenamefont {Maciejko}}]{Zerf:2016fti}%
  \BibitemOpen
  \bibfield  {author} {\bibinfo {author} {\bibfnamefont {N.}~\bibnamefont
  {Zerf}}, \bibinfo {author} {\bibfnamefont {C.-H.}\ \bibnamefont {Lin}}, \
  and\ \bibinfo {author} {\bibfnamefont {J.}~\bibnamefont {Maciejko}},\ }\href
  {\doibase 10.1103/PhysRevB.94.205106} {\bibfield  {journal} {\bibinfo
  {journal} {Phys. Rev.}\ }\textbf {\bibinfo {volume} {B94}},\ \bibinfo {pages}
  {205106} (\bibinfo {year} {2016})},\ \Eprint
  {http://arxiv.org/abs/1605.09423} {arXiv:1605.09423 [cond-mat.str-el]}
  \BibitemShut {NoStop}%
\bibitem [{\citenamefont {Mihaila}\ \emph {et~al.}(2017)\citenamefont
  {Mihaila}, \citenamefont {Zerf}, \citenamefont {Ihrig}, \citenamefont
  {Herbut},\ and\ \citenamefont {Scherer}}]{Mihaila:2017ble}%
  \BibitemOpen
  \bibfield  {author} {\bibinfo {author} {\bibfnamefont {L.~N.}\ \bibnamefont
  {Mihaila}}, \bibinfo {author} {\bibfnamefont {N.}~\bibnamefont {Zerf}},
  \bibinfo {author} {\bibfnamefont {B.}~\bibnamefont {Ihrig}}, \bibinfo
  {author} {\bibfnamefont {I.~F.}\ \bibnamefont {Herbut}}, \ and\ \bibinfo
  {author} {\bibfnamefont {M.~M.}\ \bibnamefont {Scherer}},\ }\href@noop {} {\
  (\bibinfo {year} {2017})},\ \Eprint {http://arxiv.org/abs/1703.08801}
  {arXiv:1703.08801 [cond-mat.str-el]} \BibitemShut {NoStop}%
\bibitem [{\citenamefont {Karkkainen}\ \emph {et~al.}(1994)\citenamefont
  {Karkkainen}, \citenamefont {Lacaze}, \citenamefont {Lacock},\ and\
  \citenamefont {Petersson}}]{Karkkainen:1993ef}%
  \BibitemOpen
  \bibfield  {author} {\bibinfo {author} {\bibfnamefont {L.}~\bibnamefont
  {Karkkainen}}, \bibinfo {author} {\bibfnamefont {R.}~\bibnamefont {Lacaze}},
  \bibinfo {author} {\bibfnamefont {P.}~\bibnamefont {Lacock}}, \ and\ \bibinfo
  {author} {\bibfnamefont {B.}~\bibnamefont {Petersson}},\ }\href {\doibase
  10.1016/0550-3213(94)90309-3, 10.1016/0550-3213(95)00055-W} {\bibfield
  {journal} {\bibinfo  {journal} {Nucl. Phys.}\ }\textbf {\bibinfo {volume}
  {B415}},\ \bibinfo {pages} {781} (\bibinfo {year} {1994})},\ \bibinfo {note}
  {[Erratum: Nucl. Phys.B438,650(1995)]},\ \Eprint
  {http://arxiv.org/abs/hep-lat/9310020} {arXiv:hep-lat/9310020 [hep-lat]}
  \BibitemShut {NoStop}%
\bibitem [{\citenamefont {Christofi}\ and\ \citenamefont
  {Strouthos}(2007)}]{Christofi:2006zt}%
  \BibitemOpen
  \bibfield  {author} {\bibinfo {author} {\bibfnamefont {S.}~\bibnamefont
  {Christofi}}\ and\ \bibinfo {author} {\bibfnamefont {C.}~\bibnamefont
  {Strouthos}},\ }\href {\doibase 10.1088/1126-6708/2007/05/088} {\bibfield
  {journal} {\bibinfo  {journal} {JHEP}\ }\textbf {\bibinfo {volume} {05}},\
  \bibinfo {pages} {088} (\bibinfo {year} {2007})},\ \Eprint
  {http://arxiv.org/abs/hep-lat/0612031} {arXiv:hep-lat/0612031 [hep-lat]}
  \BibitemShut {NoStop}%
\bibitem [{\citenamefont {Chandrasekharan}\ and\ \citenamefont
  {Li}(2013)}]{Chandrasekharan:2013aya}%
  \BibitemOpen
  \bibfield  {author} {\bibinfo {author} {\bibfnamefont {S.}~\bibnamefont
  {Chandrasekharan}}\ and\ \bibinfo {author} {\bibfnamefont {A.}~\bibnamefont
  {Li}},\ }\href {\doibase 10.1103/PhysRevD.88.021701} {\bibfield  {journal}
  {\bibinfo  {journal} {Phys. Rev.}\ }\textbf {\bibinfo {volume} {D88}},\
  \bibinfo {pages} {021701} (\bibinfo {year} {2013})},\ \Eprint
  {http://arxiv.org/abs/1304.7761} {arXiv:1304.7761 [hep-lat]} \BibitemShut
  {NoStop}%
\bibitem [{\citenamefont {Wang}\ \emph {et~al.}(2014)\citenamefont {Wang},
  \citenamefont {Corboz},\ and\ \citenamefont {Troyer}}]{Wang:2014cbw}%
  \BibitemOpen
  \bibfield  {author} {\bibinfo {author} {\bibfnamefont {L.}~\bibnamefont
  {Wang}}, \bibinfo {author} {\bibfnamefont {P.}~\bibnamefont {Corboz}}, \ and\
  \bibinfo {author} {\bibfnamefont {M.}~\bibnamefont {Troyer}},\ }\href
  {\doibase 10.1088/1367-2630/16/10/103008} {\bibfield  {journal} {\bibinfo
  {journal} {New J. Phys.}\ }\textbf {\bibinfo {volume} {16}},\ \bibinfo
  {pages} {103008} (\bibinfo {year} {2014})},\ \Eprint
  {http://arxiv.org/abs/1407.0029} {arXiv:1407.0029 [cond-mat.str-el]}
  \BibitemShut {NoStop}%
\bibitem [{\citenamefont {Li}\ \emph {et~al.}(2015)\citenamefont {Li},
  \citenamefont {Jiang},\ and\ \citenamefont {Yao}}]{Li:2014aoa}%
  \BibitemOpen
  \bibfield  {author} {\bibinfo {author} {\bibfnamefont {Z.-X.}\ \bibnamefont
  {Li}}, \bibinfo {author} {\bibfnamefont {Y.-F.}\ \bibnamefont {Jiang}}, \
  and\ \bibinfo {author} {\bibfnamefont {H.}~\bibnamefont {Yao}},\ }\href
  {\doibase 10.1088/1367-2630/17/8/085003} {\bibfield  {journal} {\bibinfo
  {journal} {New J. Phys.}\ }\textbf {\bibinfo {volume} {17}},\ \bibinfo
  {pages} {085003} (\bibinfo {year} {2015})},\ \Eprint
  {http://arxiv.org/abs/1411.7383} {arXiv:1411.7383 [cond-mat.str-el]}
  \BibitemShut {NoStop}%
\bibitem [{\citenamefont {Hesselmann}\ and\ \citenamefont
  {Wessel}(2016)}]{Hesselmann:2016tvh}%
  \BibitemOpen
  \bibfield  {author} {\bibinfo {author} {\bibfnamefont {S.}~\bibnamefont
  {Hesselmann}}\ and\ \bibinfo {author} {\bibfnamefont {S.}~\bibnamefont
  {Wessel}},\ }\href {\doibase 10.1103/PhysRevB.93.155157} {\bibfield
  {journal} {\bibinfo  {journal} {Phys. Rev.}\ }\textbf {\bibinfo {volume}
  {B93}},\ \bibinfo {pages} {155157} (\bibinfo {year} {2016})}\BibitemShut
  {NoStop}%
\bibitem [{\citenamefont {Chandrasekharan}\ and\ \citenamefont
  {Li}(2012)}]{Chandrasekharan:2011mn}%
  \BibitemOpen
  \bibfield  {author} {\bibinfo {author} {\bibfnamefont {S.}~\bibnamefont
  {Chandrasekharan}}\ and\ \bibinfo {author} {\bibfnamefont {A.}~\bibnamefont
  {Li}},\ }\href {\doibase 10.1103/PhysRevLett.108.140404} {\bibfield
  {journal} {\bibinfo  {journal} {Phys. Rev. Lett.}\ }\textbf {\bibinfo
  {volume} {108}},\ \bibinfo {pages} {140404} (\bibinfo {year} {2012})},\
  \Eprint {http://arxiv.org/abs/1111.7204} {arXiv:1111.7204 [hep-lat]}
  \BibitemShut {NoStop}%
\bibitem [{\citenamefont {Hands}(2016)}]{Hands:2016foa}%
  \BibitemOpen
  \bibfield  {author} {\bibinfo {author} {\bibfnamefont {S.}~\bibnamefont
  {Hands}},\ }\href {\doibase 10.1007/JHEP11(2016)015} {\bibfield  {journal}
  {\bibinfo  {journal} {JHEP}\ }\textbf {\bibinfo {volume} {11}},\ \bibinfo
  {pages} {015} (\bibinfo {year} {2016})},\ \Eprint
  {http://arxiv.org/abs/1610.04394} {arXiv:1610.04394 [hep-lat]} \BibitemShut
  {NoStop}%
\bibitem [{\citenamefont {Schmidt}\ \emph {et~al.}(2016)\citenamefont
  {Schmidt}, \citenamefont {Wellegehausen},\ and\ \citenamefont
  {Wipf}}]{Schmidt:2016rtz}%
  \BibitemOpen
  \bibfield  {author} {\bibinfo {author} {\bibfnamefont {D.}~\bibnamefont
  {Schmidt}}, \bibinfo {author} {\bibfnamefont {B.}~\bibnamefont
  {Wellegehausen}}, \ and\ \bibinfo {author} {\bibfnamefont {A.}~\bibnamefont
  {Wipf}},\ }\bibfield  {booktitle} {\emph {\bibinfo {booktitle} {{Proceedings,
  34th International Symposium on Lattice Field Theory (Lattice 2016):
  Southampton, UK, July 24-30, 2016}}},\ }\href@noop {} {\bibfield  {journal}
  {\bibinfo  {journal} {PoS}\ }\textbf {\bibinfo {volume} {LATTICE2016}},\
  \bibinfo {pages} {247} (\bibinfo {year} {2016})},\ \Eprint
  {http://arxiv.org/abs/1611.00275} {arXiv:1611.00275 [hep-lat]} \BibitemShut
  {NoStop}%
\bibitem [{\citenamefont {Rosa}\ \emph {et~al.}(2001)\citenamefont {Rosa},
  \citenamefont {Vitale},\ and\ \citenamefont {Wetterich}}]{Rosa:2000ju}%
  \BibitemOpen
  \bibfield  {author} {\bibinfo {author} {\bibfnamefont {L.}~\bibnamefont
  {Rosa}}, \bibinfo {author} {\bibfnamefont {P.}~\bibnamefont {Vitale}}, \ and\
  \bibinfo {author} {\bibfnamefont {C.}~\bibnamefont {Wetterich}},\ }\href
  {\doibase 10.1103/PhysRevLett.86.958} {\bibfield  {journal} {\bibinfo
  {journal} {Phys. Rev. Lett.}\ }\textbf {\bibinfo {volume} {86}},\ \bibinfo
  {pages} {958} (\bibinfo {year} {2001})},\ \Eprint
  {http://arxiv.org/abs/hep-th/0007093} {arXiv:hep-th/0007093 [hep-th]}
  \BibitemShut {NoStop}%
\bibitem [{\citenamefont {Hofling}\ \emph {et~al.}(2002)\citenamefont
  {Hofling}, \citenamefont {Nowak},\ and\ \citenamefont
  {Wetterich}}]{Hofling:2002hj}%
  \BibitemOpen
  \bibfield  {author} {\bibinfo {author} {\bibfnamefont {F.}~\bibnamefont
  {Hofling}}, \bibinfo {author} {\bibfnamefont {C.}~\bibnamefont {Nowak}}, \
  and\ \bibinfo {author} {\bibfnamefont {C.}~\bibnamefont {Wetterich}},\ }\href
  {\doibase 10.1103/PhysRevB.66.205111} {\bibfield  {journal} {\bibinfo
  {journal} {Phys. Rev.}\ }\textbf {\bibinfo {volume} {B66}},\ \bibinfo {pages}
  {205111} (\bibinfo {year} {2002})},\ \Eprint
  {http://arxiv.org/abs/cond-mat/0203588} {arXiv:cond-mat/0203588 [cond-mat]}
  \BibitemShut {NoStop}%
\bibitem [{\citenamefont {Braun}\ \emph {et~al.}(2011)\citenamefont {Braun},
  \citenamefont {Gies},\ and\ \citenamefont {Scherer}}]{Braun:2010tt}%
  \BibitemOpen
  \bibfield  {author} {\bibinfo {author} {\bibfnamefont {J.}~\bibnamefont
  {Braun}}, \bibinfo {author} {\bibfnamefont {H.}~\bibnamefont {Gies}}, \ and\
  \bibinfo {author} {\bibfnamefont {D.~D.}\ \bibnamefont {Scherer}},\ }\href
  {\doibase 10.1103/PhysRevD.83.085012} {\bibfield  {journal} {\bibinfo
  {journal} {Phys. Rev.}\ }\textbf {\bibinfo {volume} {D83}},\ \bibinfo {pages}
  {085012} (\bibinfo {year} {2011})},\ \Eprint {http://arxiv.org/abs/1011.1456}
  {arXiv:1011.1456 [hep-th]} \BibitemShut {NoStop}%
\bibitem [{\citenamefont {Mesterhazy}\ \emph {et~al.}(2012)\citenamefont
  {Mesterhazy}, \citenamefont {Berges},\ and\ \citenamefont {von
  Smekal}}]{Mesterhazy:2012ei}%
  \BibitemOpen
  \bibfield  {author} {\bibinfo {author} {\bibfnamefont {D.}~\bibnamefont
  {Mesterhazy}}, \bibinfo {author} {\bibfnamefont {J.}~\bibnamefont {Berges}},
  \ and\ \bibinfo {author} {\bibfnamefont {L.}~\bibnamefont {von Smekal}},\
  }\href {\doibase 10.1103/PhysRevB.86.245431} {\bibfield  {journal} {\bibinfo
  {journal} {Phys. Rev.}\ }\textbf {\bibinfo {volume} {B86}},\ \bibinfo {pages}
  {245431} (\bibinfo {year} {2012})},\ \Eprint {http://arxiv.org/abs/1207.4054}
  {arXiv:1207.4054 [cond-mat.str-el]} \BibitemShut {NoStop}%
\bibitem [{\citenamefont {Jakov\'{a}c}\ \emph {et~al.}(2015)\citenamefont
  {Jakov\'{a}c}, \citenamefont {Patk\'{o}s},\ and\ \citenamefont
  {P\'{o}sfay}}]{Jakovac:2014lqa}%
  \BibitemOpen
  \bibfield  {author} {\bibinfo {author} {\bibfnamefont {A.}~\bibnamefont
  {Jakov\'{a}c}}, \bibinfo {author} {\bibfnamefont {A.}~\bibnamefont
  {Patk\'{o}s}}, \ and\ \bibinfo {author} {\bibfnamefont {P.}~\bibnamefont
  {P\'{o}sfay}},\ }\href {\doibase 10.1140/epjc/s10052-014-3228-1} {\bibfield
  {journal} {\bibinfo  {journal} {Eur. Phys. J.}\ }\textbf {\bibinfo {volume}
  {C75}},\ \bibinfo {pages} {2} (\bibinfo {year} {2015})},\ \Eprint
  {http://arxiv.org/abs/1406.3195} {arXiv:1406.3195 [hep-th]} \BibitemShut
  {NoStop}%
\bibitem [{\citenamefont {Vacca}\ and\ \citenamefont
  {Zambelli}(2015)}]{Vacca:2015nta}%
  \BibitemOpen
  \bibfield  {author} {\bibinfo {author} {\bibfnamefont {G.~P.}\ \bibnamefont
  {Vacca}}\ and\ \bibinfo {author} {\bibfnamefont {L.}~\bibnamefont
  {Zambelli}},\ }\href {\doibase 10.1103/PhysRevD.91.125003} {\bibfield
  {journal} {\bibinfo  {journal} {Phys. Rev.}\ }\textbf {\bibinfo {volume}
  {D91}},\ \bibinfo {pages} {125003} (\bibinfo {year} {2015})},\ \Eprint
  {http://arxiv.org/abs/1503.09136} {arXiv:1503.09136 [hep-th]} \BibitemShut
  {NoStop}%
\bibitem [{\citenamefont {Borchardt}\ and\ \citenamefont
  {Knorr}(2015)}]{Borchardt:2015rxa}%
  \BibitemOpen
  \bibfield  {author} {\bibinfo {author} {\bibfnamefont {J.}~\bibnamefont
  {Borchardt}}\ and\ \bibinfo {author} {\bibfnamefont {B.}~\bibnamefont
  {Knorr}},\ }\href {\doibase 10.1103/PhysRevD.91.105011} {\bibfield  {journal}
  {\bibinfo  {journal} {Phys. Rev.}\ }\textbf {\bibinfo {volume} {D91}},\
  \bibinfo {pages} {105011} (\bibinfo {year} {2015})},\ \Eprint
  {http://arxiv.org/abs/1502.07511} {arXiv:1502.07511 [hep-th]} \BibitemShut
  {NoStop}%
\bibitem [{\citenamefont {Knorr}(2016)}]{Knorr:2016sfs}%
  \BibitemOpen
  \bibfield  {author} {\bibinfo {author} {\bibfnamefont {B.}~\bibnamefont
  {Knorr}},\ }\href {\doibase 10.1103/PhysRevB.94.245102} {\bibfield  {journal}
  {\bibinfo  {journal} {Phys. Rev.}\ }\textbf {\bibinfo {volume} {B94}},\
  \bibinfo {pages} {245102} (\bibinfo {year} {2016})},\ \Eprint
  {http://arxiv.org/abs/1609.03824} {arXiv:1609.03824 [cond-mat.str-el]}
  \BibitemShut {NoStop}%
\bibitem [{\citenamefont {Raghu}\ \emph {et~al.}(2008)\citenamefont {Raghu},
  \citenamefont {Qi}, \citenamefont {Honerkamp},\ and\ \citenamefont
  {Zhang}}]{Raghu:2007ger}%
  \BibitemOpen
  \bibfield  {author} {\bibinfo {author} {\bibfnamefont {S.}~\bibnamefont
  {Raghu}}, \bibinfo {author} {\bibfnamefont {X.-L.}\ \bibnamefont {Qi}},
  \bibinfo {author} {\bibfnamefont {C.}~\bibnamefont {Honerkamp}}, \ and\
  \bibinfo {author} {\bibfnamefont {S.-C.}\ \bibnamefont {Zhang}},\ }\href
  {\doibase 10.1103/PhysRevLett.100.156401} {\bibfield  {journal} {\bibinfo
  {journal} {Phys. Rev. Lett.}\ }\textbf {\bibinfo {volume} {100}},\ \bibinfo
  {pages} {156401} (\bibinfo {year} {2008})},\ \Eprint
  {http://arxiv.org/abs/0710.0030} {arXiv:0710.0030 [cond-mat.mes-hall]}
  \BibitemShut {NoStop}%
\bibitem [{\citenamefont {Classen}\ \emph {et~al.}(2016)\citenamefont
  {Classen}, \citenamefont {Herbut}, \citenamefont {Janssen},\ and\
  \citenamefont {Scherer}}]{Classen:2015mar}%
  \BibitemOpen
  \bibfield  {author} {\bibinfo {author} {\bibfnamefont {L.}~\bibnamefont
  {Classen}}, \bibinfo {author} {\bibfnamefont {I.~F.}\ \bibnamefont {Herbut}},
  \bibinfo {author} {\bibfnamefont {L.}~\bibnamefont {Janssen}}, \ and\
  \bibinfo {author} {\bibfnamefont {M.~M.}\ \bibnamefont {Scherer}},\ }\href
  {\doibase 10.1103/PhysRevB.93.125119} {\bibfield  {journal} {\bibinfo
  {journal} {Phys. Rev.}\ }\textbf {\bibinfo {volume} {B93}},\ \bibinfo {pages}
  {125119} (\bibinfo {year} {2016})},\ \Eprint
  {http://arxiv.org/abs/1510.09003} {arXiv:1510.09003 [cond-mat.str-el]}
  \BibitemShut {NoStop}%
\bibitem [{\citenamefont {Wetterich}(1993)}]{Wetterich:1992yh}%
  \BibitemOpen
  \bibfield  {author} {\bibinfo {author} {\bibfnamefont {C.}~\bibnamefont
  {Wetterich}},\ }\href {\doibase 10.1016/0370-2693(93)90726-X} {\bibfield
  {journal} {\bibinfo  {journal} {Phys. Lett.}\ }\textbf {\bibinfo {volume}
  {B301}},\ \bibinfo {pages} {90} (\bibinfo {year} {1993})}\BibitemShut
  {NoStop}%
\bibitem [{\citenamefont {Codello}\ \emph {et~al.}(2014)\citenamefont
  {Codello}, \citenamefont {Demmel},\ and\ \citenamefont
  {Zanusso}}]{Codello:2013bra}%
  \BibitemOpen
  \bibfield  {author} {\bibinfo {author} {\bibfnamefont {A.}~\bibnamefont
  {Codello}}, \bibinfo {author} {\bibfnamefont {M.}~\bibnamefont {Demmel}}, \
  and\ \bibinfo {author} {\bibfnamefont {O.}~\bibnamefont {Zanusso}},\ }\href
  {\doibase 10.1103/PhysRevD.90.027701} {\bibfield  {journal} {\bibinfo
  {journal} {Phys. Rev.}\ }\textbf {\bibinfo {volume} {D90}},\ \bibinfo {pages}
  {027701} (\bibinfo {year} {2014})},\ \Eprint {http://arxiv.org/abs/1310.7625}
  {arXiv:1310.7625 [hep-th]} \BibitemShut {NoStop}%
\bibitem [{\citenamefont {Codello}\ \emph {et~al.}(2017)\citenamefont
  {Codello}, \citenamefont {Safari}, \citenamefont {Vacca},\ and\ \citenamefont
  {Zanusso}}]{Codello:2017hhh}%
  \BibitemOpen
  \bibfield  {author} {\bibinfo {author} {\bibfnamefont {A.}~\bibnamefont
  {Codello}}, \bibinfo {author} {\bibfnamefont {M.}~\bibnamefont {Safari}},
  \bibinfo {author} {\bibfnamefont {G.~P.}\ \bibnamefont {Vacca}}, \ and\
  \bibinfo {author} {\bibfnamefont {O.}~\bibnamefont {Zanusso}},\ }\href@noop
  {} {\  (\bibinfo {year} {2017})},\ \Eprint {http://arxiv.org/abs/1705.05558}
  {arXiv:1705.05558 [hep-th]} \BibitemShut {NoStop}%
\bibitem [{\citenamefont {O'Dwyer}\ and\ \citenamefont
  {Osborn}(2008)}]{ODwyer:2007brp}%
  \BibitemOpen
  \bibfield  {author} {\bibinfo {author} {\bibfnamefont {J.}~\bibnamefont
  {O'Dwyer}}\ and\ \bibinfo {author} {\bibfnamefont {H.}~\bibnamefont
  {Osborn}},\ }\href {\doibase 10.1016/j.aop.2007.10.005} {\bibfield  {journal}
  {\bibinfo  {journal} {Annals Phys.}\ }\textbf {\bibinfo {volume} {323}},\
  \bibinfo {pages} {1859} (\bibinfo {year} {2008})},\ \Eprint
  {http://arxiv.org/abs/0708.2697} {arXiv:0708.2697 [hep-th]} \BibitemShut
  {NoStop}%
\bibitem [{\citenamefont {Zanusso}\ \emph {et~al.}(2010)\citenamefont
  {Zanusso}, \citenamefont {Zambelli}, \citenamefont {Vacca},\ and\
  \citenamefont {Percacci}}]{Zanusso:2009bs}%
  \BibitemOpen
  \bibfield  {author} {\bibinfo {author} {\bibfnamefont {O.}~\bibnamefont
  {Zanusso}}, \bibinfo {author} {\bibfnamefont {L.}~\bibnamefont {Zambelli}},
  \bibinfo {author} {\bibfnamefont {G.~P.}\ \bibnamefont {Vacca}}, \ and\
  \bibinfo {author} {\bibfnamefont {R.}~\bibnamefont {Percacci}},\ }\href
  {\doibase 10.1016/j.physletb.2010.04.043} {\bibfield  {journal} {\bibinfo
  {journal} {Phys. Lett.}\ }\textbf {\bibinfo {volume} {B689}},\ \bibinfo
  {pages} {90} (\bibinfo {year} {2010})},\ \Eprint
  {http://arxiv.org/abs/0904.0938} {arXiv:0904.0938 [hep-th]} \BibitemShut
  {NoStop}%
\bibitem [{\citenamefont {Vacca}\ and\ \citenamefont
  {Zanusso}(2010)}]{Vacca:2010mj}%
  \BibitemOpen
  \bibfield  {author} {\bibinfo {author} {\bibfnamefont {G.~P.}\ \bibnamefont
  {Vacca}}\ and\ \bibinfo {author} {\bibfnamefont {O.}~\bibnamefont
  {Zanusso}},\ }\href {\doibase 10.1103/PhysRevLett.105.231601} {\bibfield
  {journal} {\bibinfo  {journal} {Phys. Rev. Lett.}\ }\textbf {\bibinfo
  {volume} {105}},\ \bibinfo {pages} {231601} (\bibinfo {year} {2010})},\
  \Eprint {http://arxiv.org/abs/1009.1735} {arXiv:1009.1735 [hep-th]}
  \BibitemShut {NoStop}%
\bibitem [{\citenamefont {Zanusso}()}]{Zanusso:PhD}%
  \BibitemOpen
  \bibfield  {author} {\bibinfo {author} {\bibfnamefont {O.}~\bibnamefont
  {Zanusso}},\ }\href@noop {} {Ph.D. thesis}\BibitemShut {NoStop}%
\bibitem [{\citenamefont {Hellwig}\ \emph {et~al.}(2015)\citenamefont
  {Hellwig}, \citenamefont {Wipf},\ and\ \citenamefont
  {Zanusso}}]{Hellwig:2015woa}%
  \BibitemOpen
  \bibfield  {author} {\bibinfo {author} {\bibfnamefont {T.}~\bibnamefont
  {Hellwig}}, \bibinfo {author} {\bibfnamefont {A.}~\bibnamefont {Wipf}}, \
  and\ \bibinfo {author} {\bibfnamefont {O.}~\bibnamefont {Zanusso}},\ }\href
  {\doibase 10.1103/PhysRevD.92.085027} {\bibfield  {journal} {\bibinfo
  {journal} {Phys. Rev.}\ }\textbf {\bibinfo {volume} {D92}},\ \bibinfo {pages}
  {085027} (\bibinfo {year} {2015})},\ \Eprint
  {http://arxiv.org/abs/1508.02547} {arXiv:1508.02547 [hep-th]} \BibitemShut
  {NoStop}%
\bibitem [{\citenamefont {Synatschke}\ \emph
  {et~al.}(2009{\natexlab{a}})\citenamefont {Synatschke}, \citenamefont
  {Bergner}, \citenamefont {Gies},\ and\ \citenamefont
  {Wipf}}]{Synatschke:2008pv}%
  \BibitemOpen
  \bibfield  {author} {\bibinfo {author} {\bibfnamefont {F.}~\bibnamefont
  {Synatschke}}, \bibinfo {author} {\bibfnamefont {G.}~\bibnamefont {Bergner}},
  \bibinfo {author} {\bibfnamefont {H.}~\bibnamefont {Gies}}, \ and\ \bibinfo
  {author} {\bibfnamefont {A.}~\bibnamefont {Wipf}},\ }\href {\doibase
  10.1088/1126-6708/2009/03/028} {\bibfield  {journal} {\bibinfo  {journal}
  {JHEP}\ }\textbf {\bibinfo {volume} {03}},\ \bibinfo {pages} {028} (\bibinfo
  {year} {2009}{\natexlab{a}})},\ \Eprint {http://arxiv.org/abs/0809.4396}
  {arXiv:0809.4396 [hep-th]} \BibitemShut {NoStop}%
\bibitem [{\citenamefont {Synatschke}\ \emph
  {et~al.}(2009{\natexlab{b}})\citenamefont {Synatschke}, \citenamefont
  {Gies},\ and\ \citenamefont {Wipf}}]{Synatschke:2009nm}%
  \BibitemOpen
  \bibfield  {author} {\bibinfo {author} {\bibfnamefont {F.}~\bibnamefont
  {Synatschke}}, \bibinfo {author} {\bibfnamefont {H.}~\bibnamefont {Gies}}, \
  and\ \bibinfo {author} {\bibfnamefont {A.}~\bibnamefont {Wipf}},\ }\href
  {\doibase 10.1103/PhysRevD.80.085007} {\bibfield  {journal} {\bibinfo
  {journal} {Phys. Rev.}\ }\textbf {\bibinfo {volume} {D80}},\ \bibinfo {pages}
  {085007} (\bibinfo {year} {2009}{\natexlab{b}})},\ \Eprint
  {http://arxiv.org/abs/0907.4229} {arXiv:0907.4229 [hep-th]} \BibitemShut
  {NoStop}%
\bibitem [{\citenamefont {Mastaler}\ \emph {et~al.}(2012)\citenamefont
  {Mastaler}, \citenamefont {Synatschke-Czerwonka},\ and\ \citenamefont
  {Wipf}}]{Mastaler:2012xg}%
  \BibitemOpen
  \bibfield  {author} {\bibinfo {author} {\bibfnamefont {M.}~\bibnamefont
  {Mastaler}}, \bibinfo {author} {\bibfnamefont {F.}~\bibnamefont
  {Synatschke-Czerwonka}}, \ and\ \bibinfo {author} {\bibfnamefont
  {A.}~\bibnamefont {Wipf}},\ }\bibfield  {booktitle} {\emph {\bibinfo
  {booktitle} {{Supersymmetries and Quantum Symmetries. Proceedings, 9th
  International Workshop (SQS'2011): Dubna, Russia, July 18-23, 2011}}},\
  }\href {\doibase 10.1134/S1063779612050255} {\bibfield  {journal} {\bibinfo
  {journal} {Phys. Part. Nucl.}\ }\textbf {\bibinfo {volume} {43}},\ \bibinfo
  {pages} {593} (\bibinfo {year} {2012})}\BibitemShut {NoStop}%
\bibitem [{\citenamefont {Heilmann}\ \emph {et~al.}(2015)\citenamefont
  {Heilmann}, \citenamefont {Hellwig}, \citenamefont {Knorr}, \citenamefont
  {Ansorg},\ and\ \citenamefont {Wipf}}]{Heilmann:2014iga}%
  \BibitemOpen
  \bibfield  {author} {\bibinfo {author} {\bibfnamefont {M.}~\bibnamefont
  {Heilmann}}, \bibinfo {author} {\bibfnamefont {T.}~\bibnamefont {Hellwig}},
  \bibinfo {author} {\bibfnamefont {B.}~\bibnamefont {Knorr}}, \bibinfo
  {author} {\bibfnamefont {M.}~\bibnamefont {Ansorg}}, \ and\ \bibinfo {author}
  {\bibfnamefont {A.}~\bibnamefont {Wipf}},\ }\href {\doibase
  10.1007/JHEP02(2015)109} {\bibfield  {journal} {\bibinfo  {journal} {JHEP}\
  }\textbf {\bibinfo {volume} {02}},\ \bibinfo {pages} {109} (\bibinfo {year}
  {2015})},\ \Eprint {http://arxiv.org/abs/1409.5650} {arXiv:1409.5650
  [hep-th]} \BibitemShut {NoStop}%
\bibitem [{\citenamefont {Gies}\ and\ \citenamefont
  {Wetterich}(2002)}]{Gies:2001nw}%
  \BibitemOpen
  \bibfield  {author} {\bibinfo {author} {\bibfnamefont {H.}~\bibnamefont
  {Gies}}\ and\ \bibinfo {author} {\bibfnamefont {C.}~\bibnamefont
  {Wetterich}},\ }\href {\doibase 10.1103/PhysRevD.65.065001} {\bibfield
  {journal} {\bibinfo  {journal} {Phys. Rev.}\ }\textbf {\bibinfo {volume}
  {D65}},\ \bibinfo {pages} {065001} (\bibinfo {year} {2002})},\ \Eprint
  {http://arxiv.org/abs/hep-th/0107221} {arXiv:hep-th/0107221 [hep-th]}
  \BibitemShut {NoStop}%
\bibitem [{\citenamefont {Pawlowski}(2007)}]{Pawlowski:2005xe}%
  \BibitemOpen
  \bibfield  {author} {\bibinfo {author} {\bibfnamefont {J.~M.}\ \bibnamefont
  {Pawlowski}},\ }\href {\doibase 10.1016/j.aop.2007.01.007} {\bibfield
  {journal} {\bibinfo  {journal} {Annals Phys.}\ }\textbf {\bibinfo {volume}
  {322}},\ \bibinfo {pages} {2831} (\bibinfo {year} {2007})},\ \Eprint
  {http://arxiv.org/abs/hep-th/0512261} {arXiv:hep-th/0512261 [hep-th]}
  \BibitemShut {NoStop}%
\bibitem [{\citenamefont {Gies}(2012)}]{Gies:2006wv}%
  \BibitemOpen
  \bibfield  {author} {\bibinfo {author} {\bibfnamefont {H.}~\bibnamefont
  {Gies}},\ }\bibfield  {booktitle} {\emph {\bibinfo {booktitle} {{ECT* School
  on Renormalization Group and Effective Field Theory Approaches to Many-Body
  Systems Trento, Italy, February 27-March 10, 2006}}},\ }\href {\doibase
  10.1007/978-3-642-27320-9_6} {\bibfield  {journal} {\bibinfo  {journal}
  {Lect. Notes Phys.}\ }\textbf {\bibinfo {volume} {852}},\ \bibinfo {pages}
  {287} (\bibinfo {year} {2012})},\ \Eprint
  {http://arxiv.org/abs/hep-ph/0611146} {arXiv:hep-ph/0611146 [hep-ph]}
  \BibitemShut {NoStop}%
\bibitem [{\citenamefont {Floerchinger}\ and\ \citenamefont
  {Wetterich}(2009)}]{Floerchinger:2009uf}%
  \BibitemOpen
  \bibfield  {author} {\bibinfo {author} {\bibfnamefont {S.}~\bibnamefont
  {Floerchinger}}\ and\ \bibinfo {author} {\bibfnamefont {C.}~\bibnamefont
  {Wetterich}},\ }\href {\doibase 10.1016/j.physletb.2009.09.014} {\bibfield
  {journal} {\bibinfo  {journal} {Phys. Lett.}\ }\textbf {\bibinfo {volume}
  {B680}},\ \bibinfo {pages} {371} (\bibinfo {year} {2009})},\ \Eprint
  {http://arxiv.org/abs/0905.0915} {arXiv:0905.0915 [hep-th]} \BibitemShut
  {NoStop}%
\bibitem [{\citenamefont {Gies}\ \emph {et~al.}(2009)\citenamefont {Gies},
  \citenamefont {Synatschke},\ and\ \citenamefont {Wipf}}]{Gies:2009az}%
  \BibitemOpen
  \bibfield  {author} {\bibinfo {author} {\bibfnamefont {H.}~\bibnamefont
  {Gies}}, \bibinfo {author} {\bibfnamefont {F.}~\bibnamefont {Synatschke}}, \
  and\ \bibinfo {author} {\bibfnamefont {A.}~\bibnamefont {Wipf}},\ }\href
  {\doibase 10.1103/PhysRevD.80.101701} {\bibfield  {journal} {\bibinfo
  {journal} {Phys. Rev.}\ }\textbf {\bibinfo {volume} {D80}},\ \bibinfo {pages}
  {101701} (\bibinfo {year} {2009})},\ \Eprint {http://arxiv.org/abs/0906.5492}
  {arXiv:0906.5492 [hep-th]} \BibitemShut {NoStop}%
\bibitem [{\citenamefont {Boettcher}(2015)}]{Boettcher:2015pja}%
  \BibitemOpen
  \bibfield  {author} {\bibinfo {author} {\bibfnamefont {I.}~\bibnamefont
  {Boettcher}},\ }\href {\doibase 10.1103/PhysRevE.91.062112} {\bibfield
  {journal} {\bibinfo  {journal} {Phys. Rev.}\ }\textbf {\bibinfo {volume}
  {E91}},\ \bibinfo {pages} {062112} (\bibinfo {year} {2015})},\ \Eprint
  {http://arxiv.org/abs/1503.07817} {arXiv:1503.07817 [cond-mat.stat-mech]}
  \BibitemShut {NoStop}%
\bibitem [{\citenamefont {An}\ \emph {et~al.}(2016)\citenamefont {An},
  \citenamefont {Mesterházy},\ and\ \citenamefont {Stephanov}}]{An:2016lni}%
  \BibitemOpen
  \bibfield  {author} {\bibinfo {author} {\bibfnamefont {X.}~\bibnamefont
  {An}}, \bibinfo {author} {\bibfnamefont {D.}~\bibnamefont {Mesterházy}}, \
  and\ \bibinfo {author} {\bibfnamefont {M.~A.}\ \bibnamefont {Stephanov}},\
  }\href {\doibase 10.1007/JHEP07(2016)041} {\bibfield  {journal} {\bibinfo
  {journal} {JHEP}\ }\textbf {\bibinfo {volume} {07}},\ \bibinfo {pages} {041}
  (\bibinfo {year} {2016})},\ \Eprint {http://arxiv.org/abs/1605.06039}
  {arXiv:1605.06039 [hep-th]} \BibitemShut {NoStop}%
\bibitem [{\citenamefont {Zambelli}\ and\ \citenamefont
  {Zanusso}(2017)}]{Zambelli:2016cbw}%
  \BibitemOpen
  \bibfield  {author} {\bibinfo {author} {\bibfnamefont {L.}~\bibnamefont
  {Zambelli}}\ and\ \bibinfo {author} {\bibfnamefont {O.}~\bibnamefont
  {Zanusso}},\ }\href {\doibase 10.1103/PhysRevD.95.085001} {\bibfield
  {journal} {\bibinfo  {journal} {Phys. Rev.}\ }\textbf {\bibinfo {volume}
  {D95}},\ \bibinfo {pages} {085001} (\bibinfo {year} {2017})},\ \Eprint
  {http://arxiv.org/abs/1612.08739} {arXiv:1612.08739 [hep-th]} \BibitemShut
  {NoStop}%
\bibitem [{\citenamefont {Litim}(2000)}]{Litim:2000ci}%
  \BibitemOpen
  \bibfield  {author} {\bibinfo {author} {\bibfnamefont {D.~F.}\ \bibnamefont
  {Litim}},\ }\href {\doibase 10.1016/S0370-2693(00)00748-6} {\bibfield
  {journal} {\bibinfo  {journal} {Phys. Lett.}\ }\textbf {\bibinfo {volume}
  {B486}},\ \bibinfo {pages} {92} (\bibinfo {year} {2000})},\ \Eprint
  {http://arxiv.org/abs/hep-th/0005245} {arXiv:hep-th/0005245 [hep-th]}
  \BibitemShut {NoStop}%
\bibitem [{\citenamefont {Iliesiu}\ \emph {et~al.}(2016)\citenamefont
  {Iliesiu}, \citenamefont {Kos}, \citenamefont {Poland}, \citenamefont {Pufu},
  \citenamefont {Simmons-Duffin},\ and\ \citenamefont
  {Yacoby}}]{Iliesiu:2015qra}%
  \BibitemOpen
  \bibfield  {author} {\bibinfo {author} {\bibfnamefont {L.}~\bibnamefont
  {Iliesiu}}, \bibinfo {author} {\bibfnamefont {F.}~\bibnamefont {Kos}},
  \bibinfo {author} {\bibfnamefont {D.}~\bibnamefont {Poland}}, \bibinfo
  {author} {\bibfnamefont {S.~S.}\ \bibnamefont {Pufu}}, \bibinfo {author}
  {\bibfnamefont {D.}~\bibnamefont {Simmons-Duffin}}, \ and\ \bibinfo {author}
  {\bibfnamefont {R.}~\bibnamefont {Yacoby}},\ }\href {\doibase
  10.1007/JHEP03(2016)120} {\bibfield  {journal} {\bibinfo  {journal} {JHEP}\
  }\textbf {\bibinfo {volume} {03}},\ \bibinfo {pages} {120} (\bibinfo {year}
  {2016})},\ \Eprint {http://arxiv.org/abs/1508.00012} {arXiv:1508.00012
  [hep-th]} \BibitemShut {NoStop}%
\bibitem [{\citenamefont {Bashkirov}(2013)}]{Bashkirov:2013vya}%
  \BibitemOpen
  \bibfield  {author} {\bibinfo {author} {\bibfnamefont {D.}~\bibnamefont
  {Bashkirov}},\ }\href@noop {} {\  (\bibinfo {year} {2013})},\ \Eprint
  {http://arxiv.org/abs/1310.8255} {arXiv:1310.8255 [hep-th]} \BibitemShut
  {NoStop}%
\bibitem [{\citenamefont {Iliesiu}\ \emph {et~al.}(2017)\citenamefont
  {Iliesiu}, \citenamefont {Kos}, \citenamefont {Poland}, \citenamefont
  {Pufu},\ and\ \citenamefont {Simmons-Duffin}}]{Iliesiu:2017nrv}%
  \BibitemOpen
  \bibfield  {author} {\bibinfo {author} {\bibfnamefont {L.}~\bibnamefont
  {Iliesiu}}, \bibinfo {author} {\bibfnamefont {F.}~\bibnamefont {Kos}},
  \bibinfo {author} {\bibfnamefont {D.}~\bibnamefont {Poland}}, \bibinfo
  {author} {\bibfnamefont {S.~S.}\ \bibnamefont {Pufu}}, \ and\ \bibinfo
  {author} {\bibfnamefont {D.}~\bibnamefont {Simmons-Duffin}},\ }\href@noop {}
  {\  (\bibinfo {year} {2017})},\ \Eprint {http://arxiv.org/abs/1705.03484}
  {arXiv:1705.03484 [hep-th]} \BibitemShut {NoStop}%
\bibitem [{\citenamefont {Litim}(2001)}]{Litim:2001up}%
  \BibitemOpen
  \bibfield  {author} {\bibinfo {author} {\bibfnamefont {D.~F.}\ \bibnamefont
  {Litim}},\ }\href {\doibase 10.1103/PhysRevD.64.105007} {\bibfield  {journal}
  {\bibinfo  {journal} {Phys. Rev.}\ }\textbf {\bibinfo {volume} {D64}},\
  \bibinfo {pages} {105007} (\bibinfo {year} {2001})},\ \Eprint
  {http://arxiv.org/abs/hep-th/0103195} {arXiv:hep-th/0103195 [hep-th]}
  \BibitemShut {NoStop}%
\end{thebibliography}%

\end{document}